\documentclass[lettersize,journal]{IEEEtran}

\usepackage{amsmath,amssymb,amsfonts}
\usepackage{algorithmic}
\usepackage{algorithm}
\usepackage{array}
\usepackage[caption=false,font=normalsize,labelfont=sf,textfont=sf]{subfig}
\usepackage{textcomp}
\usepackage{stfloats}
\usepackage{url}
\usepackage{verbatim}
\usepackage{graphicx}
\usepackage{cite}
\usepackage{ifthen}
\usepackage[capitalize,nameinlink,compress]{cleveref}
\newcommand{\R}{\mathbb{R}}

\makeatletter
\newcommand{\pushright}[1]{\ifmeasuring@#1\else\omit$\displaystyle#1$\ignorespaces\fi}
\makeatother

\usepackage[prependcaption,colorinlistoftodos]{todonotes}

\newtheorem{theorem}{Theorem}
\newtheorem{lemma}{Lemma}
\newtheorem{definition}{Definition}
\newtheorem{assumption}{Assumption}

\newtheorem{problem}{Problem}
\newtheorem{remark}{Remark}

\crefname{appendix}{Appendix}{Appendices}
\crefname{figure}{Figure}{Figures}
\crefname{line}{line}{lines}
\crefname{claim}{Claim}{Claims}
\crefname{equation}{}{}
\crefname{problem}{Problem}{Problems}
\crefname{assumption}{Assumption}{Assumptions}

\newcommand{\afterReviewChangedBox}[1]{{\color{orange}#1}}

\newboolean{shortversion}
\setboolean{shortversion}{true}
\setboolean{shortversion}{false}

\begin{document}

\title{Advanced safety filter based on SOS Control Barrier and Lyapunov Functions}

\author{Michael Schneeberger, Silvia Mastellone, Florian D{\"o}rfler 
   % <-this % stops a space
   \thanks{Michael Schneeberger and Silvia Mastellone are with the Institute of Electrical Engineering, FHNW, Windisch, Switzerland (e-mails: michael.schneeberger@fhnw.ch, silvia.mastellone@fhnw.ch), and Florian D{\"o}rfler is with the Department of Information Technology and Electrical Engineering, ETH Z{\"u}rich, Switzerland, (e-mail: dorfler@control.ee.ethz.ch)} % <-this % stops a space
   \thanks{This work was supported by the Swiss National Science Foundation (SNSF) under NCCR Automation.}
}

\maketitle

\begin{abstract}
This paper presents a novel safety filter framework that ensures both safety and the preservation of the legacy control action within a nominal region.
This modular design allows the safety filter to be integrated into the control hierarchy without compromising the performance of the existing legacy controller during nominal operation. 
For a control-affine system, this is accomplished by formulating multiple Control Barrier Functions (CBFs) and Control Lyapunov-like Functions (CLFs) conditions, alongside a forward invariance condition for the legacy controller, as sum-of-squares constraints. % utilizing Putinar's Positivstellensatz.
Additionally, the state-dependent inequality constraints of the resulting Quadratic Program (QP) -- encoding the CBF and CLF conditions -- are designed to remain inactive within the nominal region, ensuring preservation of the legacy control action and performance.
\ifthenelse{\boolean{shortversion}} {} {
  Our safety filter design is also the first to include quadratic input constraints, and does not need an explicit specification of the attractor, as it is implicitly defined by the legacy controller.
}
To avoid the chattering effect and guarantee the uniqueness and Lipschitz continuity of solutions, the state-dependent inequality constraints of the QP are selected to be 
  \ifthenelse{\boolean{shortversion}} {\afterReviewChangedBox{regular}.}{regular.}
Finally, we demonstrate the method in a detailed case study involving the control of a three-phase ac/dc power converter.
\end{abstract}

% This paper presents a novel safety filter framework that ensures both safety and the preservation of the legacy control action within a nominal region. This modular design allows the safety filter to be integrated into the control hierarchy without compromising the performance of the existing legacy controller during nominal operation. For a control-affine system, this is accomplished by formulating multiple Control Barrier Functions (CBFs) and Control Lyapunov-like Functions (CLFs) conditions, alongside a forward invariance condition for the legacy controller, as sum-of-squares constraints. Additionally, the state-dependent inequality constraints of the resulting Quadratic Program (QP) -- encoding the CBF and CLF conditions -- are designed to remain inactive within the nominal region, ensuring preservation of the legacy control action and performance. To avoid the chattering effect and guarantee the uniqueness and Lipschitz continuity of solutions, the state-dependent inequality constraints of the QP are selected to be sufficiently regular. Finally, we demonstrate the method in a detailed case study involving the control of a three-phase ac/dc power converter.

\section{Introduction}

In several engineering applications, it is of much interest to endow a well-performing controller with safety guarantees by minimally changing its control action.
This concept is commonly referred to as a \textit{safety filter}, and a comprehensive overview of the subject can be found in \cite{wabersich2023data}.
An illustrative example is a finely-tuned control design based on linearization, which must be extended to adhere to hard state constraints in the nonlinear closed-loop system.
We refer to such a controller as the \emph{legacy controller}.
A popular type of safety filters for control-affine systems relies on a Control Barrier Function (CBF).
By continuously solving the associate Quadratic Program (QP), the control action of the legacy controller is smoothly adjusted as the system approaches the boundary of the \emph{safe set} $\mathcal X_s$.
The main challenges in deploying such a safety filter lies in finding viable CBF candidates and adhering to input constraints.

Even though safety filters based on a QP are well-established, they are frequently not adopted in current industrial setups due to their risk of altering the legacy control behavior during nominal operations.
In practice, the legacy control parameters are often meticulously tuned and thoroughly evaluated, which require significant time and resources.
As a result, preserving the legacy control action within the nominal operating regime is not merely a preference but a strict engineering requirement.
In such contexts, the role of a safety filter is to act as a last-resort mechanism that intervenes only when the system state approaches critical boundaries.

\begin{figure}[!t]
   \centering
   \resizebox{59mm}{!}{\includegraphics{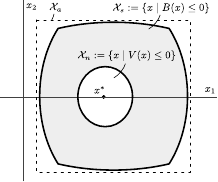}}
   \caption{
      The safe set $\mathcal X_s$, defined as the zero sublevel set of the CBF $B(x)$, specifies the set of states for which safety can be guaranteed.
      The safe set must be contained within the allowable set $\mathcal X_a$, which encodes the system's state constraints.
      Finally, the nominal region $\mathcal X_n$, defined as the zero sublevel set of the CLF $V(x)$, 
      must be contained within the safe set $\mathcal X_s$.
   }
   \label{fig:safe_and_noinal_region}
\end{figure}

This paper presents a safety filter concept that guarantees the preservation of the legacy control action within a \emph{nominal region} $\mathcal X_n$, which contains the desired closed-loop equilibrium point (or another attractor), as depicted in \cref{fig:safe_and_noinal_region}.
Unlike in a reach-avoid problem \cite{kumar2023fast}, the nominal region $\mathcal X_n$ does not represent a predetermined set of target states.
Instead, it is selected during the design process (often at an earlier stage) to encompass the nominal operating regime.
The specifications are inherently more stringent than those of a standard safety filter, but may therefore be deployed in an industrial setup without disrupting the original control architecture.

Our safety filter is realized for control-affine systems based on solving a QP at every state, where the linear inequality constraints are derived from multiple CBFs, Control Lyapunov-like Functions (CLFs), and state-dependent slack variables.
These slack variables provide local flexibility by relaxing the constraint in regions where the CBF and CLF condition are not required to hold.
A Quadratic Constraint Quadratic Program (QCQP) instead of a simple QP can be employed to encode quadratic input constraints, which are common, e.g., in power electronic applications.
Despite the restriction to control-affine and polynomial systems, our approach remains relevant for a wide range of systems encountered in practice, especially in domains like power systems and robotics, where control-affine polynomial models are commonly used.
The design process consists of two stages: first, we identify compatible CBF and CLF candidates using a Sum-of-Squares (SOS) optimization approach; second, we determine the state-dependent slack variables via a separate SOS program, ensuring that the legacy control action is preserved within the nominal region.
Our approach focuses on computing formal guarantees via SOS optimization, enabling an automated design procedure and eliminating the need for application-specific parameter tuning based on extensive simulations.

The proposed safety filter concept resembles the integration of a CLF with a CBF to attain both safety and stabilization.
Such an integration, although just with a single function, has been initially proposed in \cite{RomdlonyJayawardhana:14}, and it has since become a well-established concept, see the survey in \cite{AmesCoogan:19}.
Feasibility of such an optimization-based controller can be accomplished by introducing a slack variable that deactivates CLFs and CBFs based on their respective priorities, as shown in \cite{ames2014control}.
Alternatively, it is possible to compute compatible CLFs and CBFs using SOS tools, ensuring that for every state, there exists an input $u$ satisfying all CBF and CLF conditions.
In order to encode compatibility, there is one approach that introduces an SOS constraint for each CLF/CBF pair in the absence of inputs constraints \cite{Clark:21}.
The second approach establishes compatibility by introducing a polynomial or rational state-feedback controller, as outlined in \cite{WangHan:18,schneeberger2023sos}.
Such a controller can also be used to enforce linear input constraints, as demonstrated in \cite{WangMargellos:22}. 
Another noteworthy approach for imposing input constraints using a contrapositive statement, albeit without ensuring compatibility between CBFs and CLFs, is introduced in \cite{dai2023convex}.
These approaches introduce auxiliary polynomials that render the resulting SOS problem bilinear in its decision variables.
In \cite{TanPackard:04}, an alternating algorithm is proposed to solve this bilinear SOS problem.
One challenge in seeking CLF and CBF candidates through an alternating algorithm is to identify an initial feasible point.
In their work in \cite{zhao2023convex}, the authors constrained their search for an initial CBF to scalar functions of a specific form.
This restriction allows to reformulate the problem with linearity in its decision variables.

After our previous work \cite{schneeberger2023sos} on synthesizing compatible CLF and multiple CBFs using SOS optimization, 
we identified a recurring issue: industrial standards often requires compliance with pre-certified control behavior, particularly during nominal operations.
A prime example is the integration of a safety filter.
To address this, we proposed a modular and minimal invasive safety filter approach that preserves the legacy control behavior within the nominal region, while still ensuring safety near constraint boundaries.
To address those issues we extended the framework by including the following contributions:
\begin{itemize}
   % \item Preserving the legacy control action is achieved by ensuring that the inequality constraints of the QP remain inactive within $\mathcal X_n$.
   \item First, we synthesize compatible CLFs and CBFs based on an SOS problem, augmented by an additional SOS constraint that ensures forward-invariance of the nominal region under the legacy control.
   This constraint is essential, as it enables the construction of the proposed safety filter in a subsequent step.
   % nsures the feasibility of the proposed safety filter.}
   % enables the implementation of the proposed safety filter.
   \item Second, to increase flexibility in shaping the safe set, we employ multiple CBFs.
   Rather than imposing each CBF condition separately, we relax these conditions when the corresponding CBF is not active, similar to the idea in \cite{IsalyGhanbarpour:22}.
   Moreover, the CLF condition can also be relaxed by leveraging the forward invariance property of the CBFs, similar to \cite{valmorbida2017region} for autonomous systems.
   \item Third, given feasible CLF and CBFs candidates, we formulate a second SOS problem to compute the state-dependent slack variables that ensure safety while preserving the legacy behavior within the nominal region.
   %  to enable a safety filter implementation in which the legacy control input does not violate the inequality constrain of the QP.
   % \item Second, the desired attractor -- such as an equilibrium point or limit cycle -- stabilized by the legacy controller within the nominal region, need not be explicitly specified.
   % Instead, it is implicitly defined through a forward invariance condition imposed by the legacy controller.
   \ifthenelse{\boolean{shortversion}} {} {
     \item Fourth, we introduce a novel approach for encoding linear and quadratic input constraints on the rational controller.
     This ensures feasibility of the resulting QP-based controller, even in case of quadratic input constraints -- an issue that, to the authors' knowledge, is unsolved.
   }
\end{itemize}
\ifthenelse{\boolean{shortversion}} {} {
  To avoid chattering and guarantee the uniqueness and Lipschitz continuity of solutions, the state-dependent inequality constraints of the QP are selected to be regular (see \cref{ass:regular_safe_set}).
}
Finally, we tested our framework on a detailed three-phase ac/dc converter system connected to an infinite bus, where we successfully synthesized a safety filter that ensures satisfaction of the state constraints given by the dc-link voltage and line current, while adhering to input constraints.
% We further tracked an a priori unspecified quadrature current reference of the ac/dc converter by integrating the reference as a state into a slightly modified system model.

The remainder of the paper is structured as follows: 
In \cref{sec:preliminaries}, the preliminaries and notation are discussed.
\cref{sec:specification_and_problem_statement} introduces the concept of the advanced safety filter, its properties, and the resulting problem statement.
Numerical simulations, presented in \cref{sec:simulations}, are provided to demonstrate the efficacy of our safety filter.
Finally, \cref{sec:conclusion} concludes the paper.

\section{Preliminaries \& Notation} \label{sec:preliminaries}

\subsection{Notation}

The shorthand $\mathcal I_n := \lbrace 1, 2, ..., n \rbrace$ designates a range of natural numbers from $1$ to $n$.
$R[x]$ refers to the set of scalar polynomials in variables $x \in \R^n$, and $\Sigma[x]$ refers to the set of scalar SOS polynomials in variables $x \in \R^n$. 
If $p(x) \in \Sigma[x]$, it can be expanded to
\begin{align} \label{eq:quadratic_matrix_form}
   p(x) = Z(x)^\top Q Z(x),
\end{align}
where $Z(x)$ is a vector of monomials in $x$, and $Q$ is a square positive semidefinite matrix.
\begin{definition}
   A \emph{quadratic module} generated by polynomials $f_2, ..., f_n \in R[x]$ is defined by
   \begin{align} \label{eq:quadratic_module}
      \textbf M(f_2, ..., f_n) := \left \{ \gamma_0 + \sum_{i=2}^n \gamma_i f_i \mid \gamma_0, ..., \gamma_n \in \Sigma[x] \right \}.
   \end{align}
\end{definition}

Throughout the paper, we consider a polynomial control system, affine in the control action $u \in \R^m$, and described as
\begin{align} \label{eq:control_system}
   \dot x = f(x) + G(x) u,
\end{align}
where $x \in \R^n$ is the state, $f(x) \in \left( R[x] \right)^n$ is a polynomial vector field, and $G(x) \in \left( R[x] \right)^{n \times m}$ is a polynomial matrix.
% The control affinity property will be required to define the CBF conditions using Nagumo's theorem.
\ifthenelse{\boolean{shortversion}} {} {
  System~\cref{eq:control_system} with a polynomial state feedback control policy $u_{cl}(x) \in \left( R[x] \right)^m$ results in a closed-loop system of the form
  \begin{align} \label{eq:closed_loop_system}
     \dot x = F_{cl}(x) = f(x) + G(x) u_{cl}(x).
  \end{align}
}
The interior of a set $\mathcal X \subseteq \R^n$ is denoted by $\text{int}(\mathcal X)$.
A subset $\mathcal X \subseteq \R^n$ is called \emph{forward invariant} (cf. \cite[Theorem 4.4]{Khalil:02}) with respect to system \ifthenelse{\boolean{shortversion}} {$\dot x = f(x) + G(x) u(x)$}{\cref{eq:closed_loop_system}} if for every $x(0) \in \mathcal X$, $x(t) \in \mathcal X$ for all $t \geq 0$.
A system \ifthenelse{\boolean{shortversion}} {}{\cref{eq:closed_loop_system}} is called \emph{safe} (cf. \cite{WielandAllgower:07}) w.r.t. an \emph{allowable set} of states $\mathcal X_a \subseteq \R^n$ and the \emph{safe set} $\mathcal X_s \subseteq \R^n$, if $\mathcal X_s$ is forward invariant and $\mathcal X_s \subseteq \mathcal X_a$.
In the following, the allowable set $\mathcal X_a$ is assumed compact and regular.
\ifthenelse{\boolean{shortversion}} {} {
  We say that a controller $u_{cl}(x)$ guarantees safety if the corresponding closed-loop system is safe.

  \begin{assumption}[standing assumption] \label{ass:compact_safe_set} 
     In the following, the allowable set $\mathcal X_a$ is assumed compact and regular.
  \end{assumption}
}

\ifthenelse{\boolean{shortversion}} {} {
\subsection{Putinar's Positivstellensatz} \label{sec:putinars_psatz}

Before stating Putinar's P-satz, we need to introduce two preliminary definitions.
\begin{definition}
   A \emph{quadratic module} generated by polynomials $f_2, ..., f_n \in R[x]$ is defined by
   \begin{align} \label{eq:quadratic_module}
      \textbf M(f_2, ..., f_n) := \left \{ \gamma_0 + \sum_{i=2}^n \gamma_i f_i \mid \gamma_0, ..., \gamma_n \in \Sigma[x] \right \}.
   \end{align}
\end{definition}
\begin{definition}
   A quadratic module $\textbf M(f_2, ..., f_n)$ is said to be \emph{Archimedean} when there exists a polynomial $f \in \textbf M(f_2, ..., f_n)$ such that $\{ x \in \R^n \mid f \geq 0 \}$ is a compact set.
\end{definition}
The following theorem states Putinar's P-satz (cf. \cite[Theorem 3.20]{laurent2009sums}).
% The following theorem states Putinar's P-satz (cf. \cite{laurent2009sums}).
\begin{theorem} \label{eq:thm:putinars_psatz}
   Given polynomials $f_1, ..., f_n \in R[x]$ such that the quadratic module $\textbf M(f_2, ..., f_n)$ is Archimedean, and 
   \begin{align} \label{thm:putinars_psatz_emptyset}
      \begin{split}
          \Bigl\{ x \mid &f_1(x) \leq 0, f_2(x) \geq 0, ..., f_n(x) \geq 0 \Bigr\} = \emptyset,
      \end{split}
   \end{align}
   then 
   \begin{align} \label{thm:putinars_psatz_equality}
      f_1 \in \textbf M(f_2, ..., f_n).
   \end{align}
\end{theorem}
In words, if the set in \cref{thm:putinars_psatz_emptyset} is empty, Putinar's P-satz ensures that the polynomials $\gamma_0, \gamma_2, ..., \gamma_n$ defined in \cref{eq:quadratic_module} exist, albeit without specifying their degree.
To find these polynomials computationally requires to iteratively increase the degree of the polynomials until a solution can be found for~\cref{thm:putinars_psatz_equality}. 
An increase in the degree of the polynomials will, however, deteriorate computation time.
For many practical examples of interest, however, low-degree polynomials suffice.

Condition \cref{thm:putinars_psatz_equality} can also be expressed more explicitly through the condition that there exists $\gamma_1, \gamma_2, ..., \gamma_n \in \Sigma[x]$ such that
\begin{align} \label{eq:putinars_psatz_sos_condition}
   \gamma_1 f_1 - \gamma_2 f_2 - ... - \gamma_n f_n \in \Sigma[x].
\end{align}
Assuming the empty-set condition in \cref{thm:putinars_psatz_emptyset} holds, a solution to \cref{eq:putinars_psatz_sos_condition} involving $\gamma_1(x) = 1$ is guaranteed by Putinar's P-satz to exist when the degrees of the polynomials are possibly unbounded.
In this case, the introduction of the SOS polynomial $\gamma_1(x)$ needlessly increases the set of feasible polynomials $f_1(x)$.
However, when the degrees of the polynomials are truncated, this assertion may no longer hold true, justifying the utilization of $\gamma_1(x)$.
For the rest of the paper, we refer to both inclusions, \cref{thm:putinars_psatz_equality} and \cref{eq:putinars_psatz_sos_condition}, as an \emph{SOS constraint}.
}

\section{Specifications and Problem statement} \label{sec:specification_and_problem_statement}

\subsection{Specifications}

\begin{figure}[!t]
   \centering
   \resizebox{85mm}{!}{\includegraphics{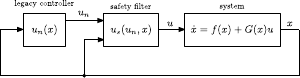}}
   \caption{
      The safety filter $u_s(u_n, x)$ adjusts the legacy control action $u_n(x)$ -- if necessary -- to guarantee safe operation.}
   \label{fig:safety_filter_diagram}
\end{figure}

The main objective of a \emph{safety filter} (see \cref{fig:safety_filter_diagram}) is to enhance the behavior of a previously designed \emph{legacy controller} $u_n(x)$ with safety guarantees.
The safety filter adjusts the control action of the legacy controller, ideally in a minimal way, to ensure the system remains within a safe set $\mathcal X_s$, which is contained within an allowable set $\mathcal X_a$ that encodes the state constraints:
% an allowable set $\mathcal X_a$, encoding the state constraints, and a safe set $\mathcal X_s$.
% The safe set denotes a set of states that is forward invariant for some state feedback controller.
% Therefore, we require that the safe set is strictly contained within the allowable set, i.e.,
\begin{align} \label{eq:xs_subsetneq_xa}
   \mathcal X_s \subseteq \mathcal X_a.
\end{align}
The safe set is preferably chosen large in volume to closely approximate the allowable set.

The standard characterization of a safety filter does not specify the extent to which it is allowed to alter the legacy control action.
This can be problematic, as the safety filter might inadvertently undermine the desired behavior of the legacy controller.
In contrast, our \emph{advanced safety filter} $u_s(x)$
specifies a distinct region $\mathcal X_n$, where the safety filter remains inactive,
that is, for all $x \in \mathcal X_n$: 
\begin{align} \label{eq:sf_inactive_condition}
   u_s(x) = u_n(x).
\end{align}
This \emph{nominal region} $\mathcal X_n$ is required to be forward invariant under the legacy control $u_n(x)$, and to be strictly contained within the safe set, i.e.,
\begin{align} \label{eq:xn_subsetneq_xs}
   \mathcal X_n \subseteq \mathcal X_s.
\end{align}
% The set inequality in \cref{eq:xn_subsetneq_xs} ensures the existence of a safety filter implementation
For a legacy controller that stabilizes the closed-loop system to an attractor within $\mathcal X_a$, such a forward invariant set does always exist.

Moreover, the advanced safety filter is designed to preserve the attractor of the legacy controller -- assuming it lies within $\mathcal X_a$. % -- across the entire safe set $\mathcal X_s$.
This property is automatically satisfied within $\mathcal X_n$ due to the preservation property \cref{eq:sf_inactive_condition} and forward invariance of $\mathcal X_n$.
Thus, it remains to ensure finite-time convergence of all trajectories originating outside $\mathcal{X}_n$ but within the safe set $\mathcal X_s$.
We define the corresponding \emph{transitional region} as:
% Moreover, the advanced safety filter should preserve such a stability guarantee across the entire safe set.
% This is already accomplished within the nominal region $\mathcal X_n$ due to its forward invariance property.
% Hence, we only need to additionally ensure a finite-time convergence to $\mathcal X_n$ for all trajectories originating within the \emph{transitional region}, defined as:
\begin{align} \label{eq:transitional_region}
   \mathcal X_t := \mathcal X_s \setminus \text{int}(\mathcal X_n),
\end{align}
where $\text{int}(\mathcal X_n)$ denotes the interior of a set $\mathcal X_n$.
\ifthenelse{\boolean{shortversion}} {} {
  \begin{remark}
     \textit{
        The volume of $\mathcal X_t$ must be balanced with the volume of $\mathcal X_n$.
        A larger $\mathcal X_t$ ensures a smooth behavior of the safety filter when returning the system to $\mathcal X_n$,
        while a larger $\mathcal X_n$ provides a wider nominal region where the safety filter does not modify the legacy control action.
     }
  \end{remark}
}

\subsection{CBF and CLF conditions formalizing the specifications} \label{sec:specifications}

\begin{figure*}[!t]
   \centering
   \setkeys{Gin}{width=0.24\linewidth}
   \subfloat[\label{fig:safety_filter_specifications1}]{\resizebox{59mm}{!}{\includegraphics{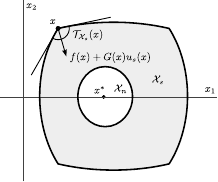}}}
   \hfil
   \subfloat[\label{fig:safety_filter_specifications2}]{\resizebox{59mm}{!}{\includegraphics{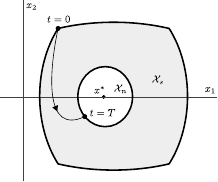}}}
   \hfil
   \subfloat[\label{fig:safety_filter_specifications3}]{\resizebox{59mm}{!}{\includegraphics{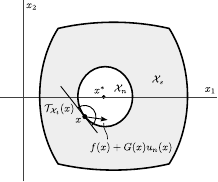}}}
   \caption{
      The advanced safety filter is characterized by the following specifications: (a) the safe set $\mathcal X_s$ is forward invariant as encoded in \cref{eq:cbf_condition}, (b) it ensures finite-time convergence towards the nominal region $\mathcal X_n$ as encoded in \cref{eq:clf_condition}, and (c) the nominal region is forward invariant under the legacy controller $u_n(x)$ as encoded in \cref{eq:forward_invariant_nominal_region} assuming $d(x) = 0$.
   }
   \label{fig:safety_filter_specifications}
\end{figure*}

We employ control Lyapunov and barrier functions to encode the specification from the previous subsection by introducing energy-like scalar functions.
In the process, we define the safe set as the intersection of zero-sublevel sets of multiple continuously differentiable functions
%  to closely fit the algebraic structure of the allowable set
\begin{align} \label{eq:allowable_set}
   \mathcal X_a := \Bigl\{ x \mid w_i(x) \leq 0 \quad \forall i \in \mathcal I_{n_{\!B}} \Bigr\} = \bigcap_{i \in \mathcal I_{n_{\!B}}} \mathcal X_{a,i},
\end{align}
where $w_i \in R[x]$ are scalar polynomials, and $\mathcal X_{a,i} := \{ x \mid w_i(x) \leq 0 \}$ denotes the corresponding zero-sublevel set.
We assume that each set $\mathcal{X}_{a,i}$ is compact, which ensures that their intersection $\mathcal{X}_a$ is also compact.
This compactness condition is important for the derivation of the SOS constraints, and it is well aligned with our intended application scenarios, where state constraints are typically bounded.
% This compactness condition is important for applying Putinar's P-satz in the next section, and it is well aligned with our intended application scenarios, where state constraints are typically bounded.

The forward invariance property of the safe set can be asserted by a scalar function $B(x)$ defined as the point-wise maximum of a set of continuously differentiable functions $B_i: \R^n \to \R$ indexed by $\mathcal I_{n_{\!B}}$:
\begin{align} \label{eq:control_barrier_function}
   B(x) := \underset{i \in \mathcal I_{n_{\!B}}}{\text{max}} \, B_i(x),
\end{align}
such that the safe set is given by the intersection of the individual zero-sublevel sets:
\begin{align*}
   \mathcal X_s := \left \{ x \mid B(x) \leq 0 \right \} = \bigcap_{i \in \mathcal I_{n_{\!B}}} \mathcal X_{s,i},
\end{align*}
where $\mathcal X_{s,i} := \left \{ x \mid B_i(x) \leq 0 \right \}$.
This multi-function construction provides greater flexibility in shaping the safe set compared to relying on a single barrier function.
In particular, each function $B_i(x)$ is allowed to define a sublevel set that may partially extend beyond the allowable set, as long as their intersection $\mathcal X_s$ remains within the allowable set $\mathcal X_a$. 
By contrast, a single barrier function must define a sublevel set that is entirely contained within the allowable set, which leads to more conservative safe sets.
Hence, to enforce the containment condition in \cref{eq:xs_subsetneq_xa}, we impose the subset relation
\begin{align} \label{eq:xs_subsetneq_xa_i}
   \mathcal X_{s,i} \subseteq \mathcal X_{a,i},
\end{align}
for every $i \in \mathcal I_{n_{\!B}}$.
% This can be seen from $\bigcap_{i \in \mathcal I_{n_{\!B}}} \mathcal X_{s,i} \subseteq \bigcap_{i \in \mathcal I_{n_{\!B}}} \mathcal X_{a,i}$ implying $\mathcal X_s \subseteq \mathcal X_a$.
This condition ensures that the overall safe set $\mathcal X_s = \bigcap_{i \in \mathcal I_{n_{\!B}}} \mathcal X_{s,i}$ is contained within the allowable set $\mathcal X_a =  \bigcap_{i \in \mathcal I_{n_{\!B}}} \mathcal X_{a,i}$.
We assume that the barrier function $B(x)$ is constructed from the same number of components $B_i(x)$ as the number of polynomials $w_i(x)$ that define the allowable set.
This condition is not restrictive in practice, since one can define multiple $B_i(x)$ functions to share the same expression, effectively reducing the number of distinct subsets $\mathcal X_{s,i}$ compared to the subsets $\mathcal X_{a,i}$.
Conversely, it is also possible to define additional distinct $B_i(x)$ functions even if some $w_i(x)$ are repeated.
% We assume that the CBF $B(x)$ is defined using the same number of polynomials $B_i(x)$ as polynomials $w_i(x)$ that define the allowable set.
% To reduce the number of decision variables in $B(x)$, we can equate certain polynomials $B_i(x)$ in the resulting SOS problem.
% This reduction results in a safe set comprising fewer distinct subsets $\mathcal X_{s,i}$ compared to the subsets $\mathcal X_{a,i}$ that define the allowable set.}
\begin{assumption} \label{ass:regular_safe_set}
   We assume the safe set $\mathcal X_s$ to be regular in the sense that $\forall x \in \mathcal X_s, \exists z \in \R^n, \forall i \in \mathcal I_{n_{\!B}}$ \cite[Def. 4.9]{Blanchini:99}:
   \begin{align*}
      B_i(x) + \nabla B_i(x)^\top z < 0
   \end{align*}
\end{assumption}
The regularity assumption in \cref{ass:regular_safe_set} is a mild one, and it generically holds true, 
if $B_i(x)$ is chosen as a polynomial.
It implies that the interior of the safe set is given by $\text{int} (\mathcal X_s) = \left \{ x \mid B_i(x) < 0 \,\, \forall i \in \mathcal I_{n_{\!B}} \right \}$.
Examples of polynomials that violate the regularity assumption can be found in \cite[Chapter 4.2]{Blanchini:99}.

To ensure forward invariance of $\mathcal X_s$, as shown in \cref{fig:safety_filter_specifications}(a), $B(x)$ needs to satisfy Nagumo's theorem, that is, $\forall x \in \partial \mathcal X_s, \exists u \in \R^m$ (cf. \cite[Nagumo's Theorem 4.7]{Blanchini:99}):
\begin{align} \label{eq:cbf_condition}
   f(x) + G(x) u \in \mathcal T_{\mathcal X_s}(x),
\end{align}
where, under \cref{ass:regular_safe_set}, the tangent cone is defined by
\begin{align} \label{eq:cbf_condition_active_constraints}
   \mathcal T_{\mathcal X_s}(x) := \{ z \mid \nabla B_i(x)^\top z \leq 0 \quad i \in \text{Act}(x) \},
\end{align}
and $\text{Act}(x) := \left\{ i \in \mathcal I_{n_{\!B}} \mid B_i(x) = 0 \right\}$ denotes the set of active constraints.
The function $B(x)$ in \cref{eq:control_barrier_function} satisfying \cref{eq:cbf_condition,eq:cbf_condition_active_constraints} will be referred to as a \emph{Control Barrier Function} (CBF).

Compared to the standard safety filter, our approach additionally require finite-time convergence to the nominal region $\mathcal X_n$. %, as outlined in the specifications.
This is enforced via a Control Lyapunov Function (CLF), defined by a continuously differentiable function $V: \R^n \to \R$.
In the transitional region $\mathcal X_t$, this function needs to satisfy a Control Lyapunov-like condition, as illustrated in \cref{fig:safety_filter_specifications}(b).
Namely, $\forall x \in \mathcal X_t, \exists u \in \R^m$:
\begin{align} \label{eq:clf_condition}
   \nabla V(x)^\top \left ( f(x) + G(x) u \right ) + d(x) \leq 0
\end{align}
with \emph{dissipation rate}
\begin{align} \label{eq:dissipation_rate}
   d(x) := \lambda(x) \left ( V(x) + V_0 \right ) > 0,
\end{align}
where $\lambda: \R^n \to \R_{> 0}$ is a Lipschitz continuous function, and $V_0 \leq \text{min}_x \left ( V(x) \right )$ is a constant value, ensuring $d(x)$ is strictly positive (and thus $V(x)$ strictly decaying) in the transitional region $\mathcal X_t$.
By utilizing a strictly positive function $d(x)$, the CLF condition can be expressed as a non-strict inequality condition similar to the CBF condition -- an important distinction for deriving the SOS constraints.
We define the nominal region as the zero sublevel set of $V(x)$:
\begin{align*}
   \mathcal X_n := \left \{ x \mid V(x) \leq 0 \right \}.
\end{align*}
Consequently, the transitional region is given by:
\begin{align*}
   \mathcal X_t := \left \{ x \mid B(x) \leq 0 \leq V(x) \right \}.
\end{align*}
The following lemma establishes the finite-time convergence property induced by the CLF $V(x)$. 
A detailed proof is provided in \ifthenelse{\boolean{shortversion}} {
  \cite{schneeberger2024advanced}
}{
  \cref{sec:proof_finite_time_convergence}
}.
\begin{lemma} \label{thm:finite_time_convergence}
   Given a Lipschitz-continuous controller $u_s: \R^n \to \R^m$ that satisfies condition \cref{eq:clf_condition} with $u := u_s(x)$ and renders the safe set $\mathcal X_s$ forward invariant,
   then for all trajectories of system $\dot x = f(x) + G(x) u_s(x)$ starting in the safe set $\mathcal X_s$ there exists some finite $T \geq 0$ such that $x(t) \in \mathcal X_n$ for all $t \geq T$.
\end{lemma}

Additionally, to make the nominal region forward invariant w.r.t. the legacy controller $u_n(x)$, as depicted in \cref{fig:safety_filter_specifications}(c), we propose the condition that $\forall x \in \partial \mathcal X_n$:
\begin{align} \label{eq:forward_invariant_nominal_region}
   \nabla V(x)^\top \left ( f(x) + G(x) u_n(x) \right ) + d(x) \leq 0.
\end{align}
Notice that forward invariance also follows with $d(x) = 0$.
By adopting a slightly stricter forward invariance condition in \cref{eq:forward_invariant_nominal_region} involving the strictly positive dissipation rate $d(x)$, we are able to proof the existence of a safety filter implementation in \cref{sec:online_controller}.
\begin{remark}
   \textit{
      Compared to approaches that simultaneously establish safety and stability \cite{RomdlonyJayawardhana:14}, our proposed specifications are less restrictive.
      % In fact, feasible CBF and CLF candidates that satisfy these specifications can always be found, provided that a forward invariant set $\mathcal X_n$ under the legacy controller exists that satisfies \cref{eq:forward_invariant_nominal_region} and 
      In fact, feasible CBF and CLF candidates that satisfy these specifications can always be found, provided that a forward invariant set under the legacy controller exists -- including a dissipation term as in \cref{eq:forward_invariant_nominal_region} -- that is fully contained within the allowable set.
      To such cases, a trivial but valid construction would be to set all barrier and Lyapunov functions equal, such that CBF and CLF condition \cref{eq:cbf_condition,eq:clf_condition} are automatically satisfied by the forward invariance condition of the legacy controller.
      % , set $V(x) = B_1(x) = \hdots = B_t(x)$ such that CBF and CLF condition \cref{eq:cbf_condition,eq:clf_condition} are automatically satisfied by the forward invariance condition \cref{eq:forward_invariant_nominal_region} of the legacy controller.
      % As the volume of the transitional region $\mathcal X_t$ approaches zero, the proposed safety filter transitions into a safety filter based on a single CBF $V(x) = B_1(x) = \hdots = B_t(x)$, where the CBF and CLF condition \cref{eq:cbf_condition,eq:clf_condition} are automatically satisfied by the forward invariance condition \cref{eq:forward_invariant_nominal_region} of the legacy controller.
      % Therefore, feasible CBF and CLF candidates can be found (in the limit) if a forward invariant set for the system, governed by the legacy controller, exists and is fully contained within the allowable set.
      %
      % the problem simplifies to finding a forward invariant set for the system, using the legacy controller, that is contained in the allowable set.
      % In this case, a solution to the proposed safety filter exists if there exists a forward invariant set using the legacy controller contained within the allowable set.
   }
\end{remark}

\subsection{Problem statement} \label{sec:problem_statement}

The advanced safety filter can be divided into two subproblems.
The first resolves around finding $B(x)$ and $V(x)$ that fulfill the CBF and CLF condition outlined in the preceding section.
% To verify the mutual feasibility of these conditions -- specifically \cref{eq:cbf_condition,eq:clf_condition} -- we introduce a rational feedback controller $u_\text{SOS}(x)$, which serves as a candidate input that simultaneously satisfies both inequalities.
% The challenge of deriving the advanced safety filter can be divided into two subproblems.
% The first subproblem resolves around finding $B(x)$ and $V(x)$ that fulfill the CBF and CLF condition outlined in the preceding section.
% This involves constructing a rational controller $u_\text{SOS}(x)$ to prove the compatibility of the CLF and CBF conditions in \cref{eq:cbf_condition,eq:clf_condition}.
% Furthermore, we necessitate the assurance of the forward invariance condition of the legacy controller as defined in \cref{eq:forward_invariant_nominal_region}.
\begin{problem} \label{prob:finding_cbfs_and_clf}
   Given a dynamical system \cref{eq:control_system}, an allowable set of states $\mathcal X_a$ and a legacy controller $u_n(x) \in \mathcal U$, find scalar functions $B(x): \R^n \to \R$ and $V(x): \R^n \to \R$ such that: \begin{itemize}
      \item (set containment) the containment conditions in \cref{eq:xs_subsetneq_xa,eq:xn_subsetneq_xs} are satisfied,
      \item (finite-time convergence) the CLF condition in \cref{eq:clf_condition} is satisfied,
      \item (forward-invariance of $\mathcal X_s$) CBF and CLF conditions in \cref{eq:cbf_condition,eq:clf_condition} are satisfied on $\partial \mathcal X_s$ for the same $u$,
      \item (forward-invariance of $\mathcal X_n$) legacy control condition in \cref{eq:forward_invariant_nominal_region} is satisfied,
      \item (objective) CBF $B(x)$ maximizes the volume of the safe set, denoted by $\text{vol}(\mathcal X_s)$.
   \end{itemize}
\end{problem}

For the second subproblem, a safety filter $u_s(x)$ is constructed based on the CBF and CLF.
% previously determined functions $B(x)$ and $V(x)$.
% This controller must be feasible and Lipschitz-continuous to ensure a smooth transitioning from the safety filter operation to the legacy control operation, and must enforce the CBF and CLF conditions while remaining inactive on the nominal region $\mathcal X_n$.
\begin{problem} \label{prob:finding_safety_filter}
   Given a dynamical system \cref{eq:control_system}, a Lipschitz-continuous legacy controller $u_n(x)$, a CBF $B(x)$ and CLF $V(x)$, as specified in \cref{prob:finding_cbfs_and_clf},
   find state-feedback control law $u_s(x)$ that:
   \begin{itemize}
      \item is feasible for all $x \in \R^n$,
      \item satisfies the CBF and CLF conditions \cref{eq:cbf_condition,eq:clf_condition} with $u := u_s(x)$,
      \item coincides with $u_n(x)$ on $\mathcal X_n$ (cf. \cref{eq:sf_inactive_condition}),
      \item is Lipschitz-continuous.
   \end{itemize}
\end{problem}
% For the second subproblem, a safety filter $u_s(x)$  is constructed from $B(x)$ and $V(x)$, which is guaranteed to be feasible and Lipschitz-continuous, entailing a smooth transitioning from the safety filter operation to the legacy control operation.
% \begin{problem} \label{prob:finding_safety_filter}
%    Given a dynamical system \cref{eq:closed_loop_system}, a Lipschitz-continuous legacy controller $u_n(x)$, a CBF $B(x)$ and CLF $V(x)$, as specified in \cref{prob:finding_cbfs_and_clf},
%    find state-feedback control law $u_s(x)$ that
%    is feasible for all $x \in \R^n$, satisfies the CBF and CLF conditions \cref{eq:cbf_condition,eq:clf_condition} with $u := u_s(x)$, coincides with $u_n(x)$ on $\mathcal X_n$ (cf. \cref{eq:sf_inactive_condition}), and is Lipschitz-continuous.
% \end{problem}
In \cref{sec:sos_optimization}, we delve into a solution for \cref{prob:finding_cbfs_and_clf}, while a solution for \cref{prob:finding_safety_filter} is derived in \cref{sec:online_controller}.

\begin{figure}[!t]
   \centering
   \resizebox{85mm}{!}{\includegraphics{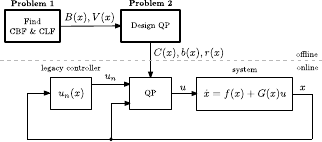}}
   \caption{
      \cref{prob:finding_cbfs_and_clf} involves searching for compatible CBF $B(x)$ and CLF $V(x)$ using SOS tools.
      Based on $B(x)$ and $V(x)$, the linear constraints of the QP given by $C(x)$, $b(x)$ and $r(x)$ are determined.}
\end{figure}

\section{Constructing CBF and CLF using SOS optimization} \label{sec:sos_optimization}

\subsection{SOS formulation of CBF and CLF conditions} \label{sec:cbf_and_clf_sos_constraints}

\ifthenelse{\boolean{shortversion}} {
  Finding $B(x)$ and $V(x)$ that fulfill the CBF and CLF condition outlined in the preceding section involves the construction of a rational controller $u_\text{SOS}(x) := p(x) / s(x)$ to prove the compatibility of the CLF and CBF conditions in \cref{eq:cbf_condition,eq:clf_condition}.
  The last SOS constraint ensures the forward invariance condition of the legacy controller, as defined in \cref{eq:forward_invariant_nominal_region}.
  % Furthermore, we necessitate the assurance of the forward invariance condition of the legacy controller as defined in \cref{eq:forward_invariant_nominal_region}.
  The resulting SOS constraints are summarized as:
  \begin{subequations} \label{eq:sos_constr}
     \begin{align}
        \begin{split} \label{eq:sos_constr_cbf}
           -\nabla B_i^\top \left ( s f + G p \right ) &\in \textbf M(B_i, -B_{\mathcal I_{n_{\!B}}})
        \end{split} \\
        \begin{split} \label{eq:sos_constr_clf}
           -\nabla V^\top \left ( s f + G p \right ) - s d &\in \textbf M(V, -B_{\mathcal I_{n_{\!B}}})
        \end{split} \\
        \begin{split} \label{eq:sos_constr_nom}
           -\nabla V^\top \left ( f + G u_n \right ) - d &\in \textbf M(V, -V),
        \end{split}
     \end{align}
  \end{subequations}
  for $i \in \mathcal I_{n_{\!B}}$, where $B_{\mathcal I_t}(x) = \begin{bmatrix}B_1(x) & \hdots & B_t(x)\end{bmatrix}^\top$ is a vector of CBFs, $p \in (R[x])^m$ is a polynomial vector, and $s \in \Sigma[x]$ is a scalar polynomial.
  The full derivation of the SOS constraint can be found in \cite{schneeberger2024advanced}.
}{
  The objective of this subsection is to derive sufficient SOS conditions that encode the CBF condition on $\mathcal X_s$ and $\mathcal X_n$, as specified in \cref{eq:cbf_condition,eq:forward_invariant_nominal_region}, and the CLF condition, as specified in \cref{eq:clf_condition}.
  We further demonstrate that these SOS conditions are not overly conservative and pose no practical concerns.

  The CBF condition in \cref{eq:cbf_condition} involves the set of active constraints $\text{Act}(x)$, a construction that cannot be directly expressed as an SOS constraint.
  To address this issue, we first consider the polynomials $B_i(x)$, $i \in \mathcal I_t$, defining the CBF $B(x)$ in \cref{eq:control_barrier_function}, as independent CBFs in their own right.
  By ensuring compatibility among these CBFs $B_i(x)$ as proposed in \cite{schneeberger2023sos}, we demonstrate sufficiency of the resulting SOS constraints with respect to CBF condition in \cref{eq:cbf_condition} (involving the desired CBF $B(x)$).
  The CBF condition for each $B_i(x)$, $i \in \mathcal I_t$, is expressed as $\forall x \in \{ x \in \mathcal X_s \mid B_i(x) = 0\}, \exists u \in \R^m$:
  \begin{align} \label{eq:cbf_condition_inequality}
     \nabla B_i(x)^\top \left ( f(x) + G(x) u \right ) \leq 0.
  \end{align}

  Another challenge in converting conditions \cref{eq:cbf_condition_inequality,eq:clf_condition} to SOS constraints is the presence of the existential quantifier $\exists u \in \R^m$.
  However, following the approach in \cite{TanPackard:04}, we can reformulate these conditions without the need for an existential quantifier.
  The CBF condition \cref{eq:cbf_condition_inequality} can be expressed as $\forall x \in \{ x \in \mathcal X_s \mid B_i(x) = 0, \nabla B_i(x)^\top G(x) = 0 \}$:
  \begin{align} \label{eq:cbf_condition_no_exist_quant}
     \nabla B_i(x)^\top f(x) \leq 0.
  \end{align}
  Similarly, the CLF condition in \cref{eq:clf_condition} can be formulated as $\forall x \in \{ x \in \mathcal X_t \mid \nabla V(x)^\top G(x) = 0 \}$:
  \begin{align} \label{eq:clf_condition_no_exist_quant}
     \nabla V(x)^\top f(x) + d(x) \leq 0.
  \end{align}

  In order to render the problem tractable, we restrict our attention to polynomial candidates $B_i \in R[x]$, $i \in \mathcal I_t$, and polynomial candidate $V \in R[x]$.
  We then apply Putinar's P-satz, as stated in \cref{eq:thm:putinars_psatz}, on a slightly more stringent conditions compared to \cref{eq:forward_invariant_nominal_region,eq:cbf_condition_no_exist_quant,eq:clf_condition_no_exist_quant}, where the inequality is replaced by a strict inequality.
  These more stringent conditions are summarized as the following empty-set conditions, $i \in \mathcal I_t$:
  \begin{subequations} \label{eq:empty-set_conditions}
     \begin{align}
        \begin{split}
           \{ x \mid &-\nabla B_i(x)^\top f(x) \leq 0, \nabla B_i(x)^\top G(x) = 0, \\
           &B_i(x) = 0, -B_{\mathcal I_t}(x) \geq 0 \} = \emptyset
        \end{split} \\
        \begin{split} \label{eq:empty-set_conditions_clf}
           \{ x \mid &- \nabla V(x)^\top f(x) - d(x) \leq 0, \nabla V(x)^\top G(x) = 0, \\
           & V(x) \geq 0, -B_{\mathcal I_t}(x) \geq 0 \} = \emptyset
        \end{split} \\
        \begin{split}
           \{ x \mid &\nabla V(x)^\top \left ( f(x) + G(x) u_n(x) \right ) + d(x) \leq 0, \\
           &V(x) = 0 \} = \emptyset,
        \end{split}
     \end{align}
  \end{subequations}
  where $B_{\mathcal I_t}(x) = \begin{bmatrix}B_1(x) & \hdots & B_t(x)\end{bmatrix}^\top$ is a vector of CBFs.
  The resulting SOS constraints are summarized below.
  A detailed derivation of the SOS constraint \cref{eq:sos_constraint_incompatible} from the empty-set condition \cref{eq:empty-set_conditions} is given in \cref{sec:derive_sos_constr}.
  \begin{subequations} \label{eq:sos_constraint_incompatible}
     \begin{align}
        \begin{split} \label{eq:sos_constraint_incompatible_cbf}
           -\nabla B_i^T \left ( s_i f + G p_i \right ) &\in \textbf M(B_i, -B_{\mathcal I_t})
        \end{split} \\
        \begin{split} \label{eq:sos_constraint_incompatible_clf}
           -\nabla V^T \left ( s_0 f + G p_0 \right ) - s_0 d &\in \textbf M(V, -B_{\mathcal I_t})
        \end{split} \\
        \begin{split} \label{eq:sos_constraint_incompatible_cbf_un}
           -\nabla V^T \left ( f + G u_n \right ) - d &\in \textbf M(V, -V),
        \end{split}
     \end{align}
  \end{subequations}
  for some $p_i \in (R[x])^m$ and $s_i \in \Sigma[x]$, $i \in \{ 0 \} \cup \mathcal I_t$.
  For clarity, we omit the brackets and variables in the function notation.
  To achieve the simultaneous satisfaction of CBF and CLF conditions \cref{eq:cbf_condition,eq:clf_condition} with a single input $u := p(x) / s(x)$ on $\partial \mathcal X_s$, as specified in \cref{prob:finding_cbfs_and_clf}, we equate the polynomials $p_i(x)$ and $s_i(x)$, $i \in \{ 0 \} \cup \mathcal I_t$.
  The resulting SOS constraints are given as:
  \begin{subequations} \label{eq:sos_constr}
     \begin{align}
        \begin{split} \label{eq:sos_constr_cbf}
           -\nabla B_i^T \left ( s f + G p \right ) &\in \textbf M(B_i, -B_{\mathcal I_t})
        \end{split} \\
        \begin{split} \label{eq:sos_constr_clf}
           -\nabla V^T \left ( s f + G p \right ) - s d &\in \textbf M(V, -B_{\mathcal I_t})
        \end{split} \\
        \begin{split} \label{eq:sos_constr_nom}
           -\nabla V^T \left ( f + G u_n \right ) - d &\in \textbf M(V, -V),
        \end{split}
     \end{align}
  \end{subequations}
  for $i \in \mathcal I_t$, and some $p \in (R[x])^m$, and $s \in \Sigma[x]$.

  \begin{remark}
     \textit{
        The polynomials $p(x)$ and $s(x)$ induce (after dividing \cref{eq:sos_constr_cbf} and \cref{eq:sos_constr_clf} by $s(x)$) a rational controller $u_\text{SOS}(x) := p(x) / s(x)$, employed to prove compatibility of the CBFs $B_i(x)$ and CLF $V(x)$.
     }
  \end{remark}

  The proof of \cref{lem:sos_constraint_sufficient} can be found in \cref{sec:proof_sos_constraint_sufficient}.
  \begin{lemma} \label{lem:sos_constraint_sufficient}
     If $s(x) > 0$ and \cref{ass:regular_safe_set} is satisfied, then the SOS constraints in \cref{eq:sos_constr} are sufficient conditions for the CBF and CLF property in \cref{eq:cbf_condition,eq:clf_condition,eq:forward_invariant_nominal_region} to hold simultaneously with $u := p(x) / s(x)$.
  \end{lemma}

  The SOS constraints in \cref{eq:sos_constr} imply the inequalities since $p\in \Sigma[x]$ implies $p(x) \geq 0$:
  \begin{subequations} \label{eq:sos_constraint_inequality}
     \begin{align}
        \begin{split} \label{eq:sos_constraint_inequality_cbf}
           \nabla B_i^\top \left ( s f + G p \right ) + \gamma_{0,i} B_i - \gamma_{1,i}^\top B_{\mathcal I_t} &\leq 0 
        \end{split} \\
        \begin{split} \label{eq:sos_constraint_inequality_clf}
           \nabla V^\top \left ( s f + G p \right ) + s d + \gamma_{0,0} V - \gamma_{1,0}^\top B_{\mathcal I_t} &\leq 0
        \end{split} \\
        \begin{split} \label{eq:sos_constraint_inequality_nom}
           \nabla V^\top \left ( f + G u_n \right ) + d + \gamma_n V &\leq 0,
        \end{split}
     \end{align}
  \end{subequations}
  for some $\gamma_{0,i}, \gamma_n \in R[x]$, $p \in (R[x])^m$, $s, \gamma_{0,0} \in \Sigma[x]$, and $\gamma_{1,i}, \gamma_{1,0} \in (\Sigma[x])^\top$, $i \in \mathcal I_t$.

  \cref{lem:sos_constraint_sufficient} only provides sufficient conditions.
  However, adding an arbitrary small perturbation $\epsilon > 0$ to the SOS constraints in \cref{eq:sos_constr} results in necessary conditions when restricting to rational controllers $u_\text{SOS}(x) = p(x) / s(x)$.
  This is because \cref{eq:cbf_condition,eq:clf_condition,eq:forward_invariant_nominal_region} with $u := u_\text{SOS}(x)$ imply strict inequalities involving $\epsilon$ as follows: $p(x) \leq 0$ implies $p(x) - \epsilon < 0$.
  These strict inequalities imply the SOS constraints \cref{eq:sos_constr} perturbed by $\epsilon$ according to \cref{eq:thm:putinars_psatz}.
  Hence, for all practical purposes, the solution set of the SOS constraints in \cref{eq:sos_constr} encompasses the desired set of CBFs and CLFs.

  \subsection{SOS formulation of input constraints} \label{sec:input_sos_constraints}

  In the following, we consider two types of inputs constraints: linear and quadratic input constraints.
  Consider the linear input constraint set
  \begin{align} \label{eq:linear_input_constraints}
     \mathcal U = \{ u \mid A u \leq b \},
  \end{align}
  where $A \in \R^{m \times l}$ and $b \in \R^l$.
  These linear input constraints can be encoded by the SOS constraints:
  \begin{align} \label{eq:linear_input_sos_constraints}
     \begin{split}
        -a_i p(x) + b_i s(x) \in \textbf M(-B_{\mathcal I_t}),
     \end{split}
  \end{align}
  where $a_i$ is the $i$th row of $A$, $i \in \mathcal I_l$, and $p(x)$ and $s(x)$ are as in \cref{eq:sos_constr}.
  Next, we consider the quadratic input constraint
  \begin{align} \label{eq:quadratic_input_constraints}
     \mathcal U = \{ u \mid (u - u_0)^\top Q_u (u - u_0) \leq u_\text{max}^2 \},
  \end{align}
  for some $Q_u \succeq 0$ and $u_0 \in \R^m$.
  The quadratic input constraint \cref{eq:quadratic_input_constraints} can be encoded using the SOS constraint involving variables $x \in \R^n$ and $v \in \R^m$:
  \begin{align} \label{eq:quadratic_input_sos_constraints}
     \begin{split}
        &- v^\top Q_u \left ( p(x) - s(x) u_0 \right ) + s(x) u_\text{max}^2 \in \textbf M(-f_u, -B_{\mathcal I_t}),
     \end{split}
  \end{align}
  where $f_u(x, v) := v^\top Q_u v - u_\text{max}^2 \in R[x,v]$ encodes the input constraint set in \cref{eq:quadratic_input_constraints}.
  The following lemma establishes the implication of the input constraints based on the SOS constraints.
  The proof is given in \cref{sec:proof_quadratic_input}.
  \begin{lemma} \label{lem:quadratic_input_sos_constraints}
     If $s(x) > 0$, a solution $p(x)$ and $s(x)$ to the SOS constraint \cref{eq:linear_input_sos_constraints}, resp. \cref{eq:quadratic_input_sos_constraints}, also satisfies the input constraints \cref{eq:linear_input_constraints}, resp. \cref{eq:quadratic_input_constraints}, with $p(x)/s(x) \in \mathcal U$ for all $x \in \mathcal X_s$.
  \end{lemma}

  \subsection{Safe set and nominal region requirement}

  In this section, we express the containment conditions of the allowable set $\mathcal X_a$, the safe set $\mathcal X_s$ and the nominal region $\mathcal X_n$, as defined in \cref{eq:xs_subsetneq_xa_i,eq:xn_subsetneq_xs} (cf. \cref{fig:safe_and_noinal_region}), in terms of SOS constraints.
  A slightly more stringent condition compared to \cref{eq:xs_subsetneq_xa_i}, can be expressed as the empty-set condition for all $i \in \mathcal I_t$:
  \begin{align} \label{eq:xs_subsetneq_xa_emptyset}
     \{ x \mid B_i(x) \leq 0, w_i(x) \geq 0 \} = \emptyset.
  \end{align}
  Using Putinar's P-satz in \cref{eq:thm:putinars_psatz}, we translate \cref{eq:xs_subsetneq_xa_emptyset} into the SOS constraints for all $i \in \mathcal I_t$:
  \begin{align} \label{eq:sos_constraint_xs_subsetneq_xa}
     B_i \in \textbf M(w_i).
  \end{align}
  Similarly, the containment condition of $\mathcal X_n$ inside $\mathcal X_s$ in \cref{eq:xn_subsetneq_xs}, can be expressed as the empty-set condition for all $i \in \mathcal I_t$:
  \begin{align} \label{eq:xn_subsetneq_xs_emptyset}
     \{ x \mid V(x) \leq 0, B_i(x) \geq 0 \} = \emptyset.
  \end{align}
  Using Putinar's P-satz, we translate \cref{eq:xn_subsetneq_xs_emptyset} into the SOS constraints for all $i \in \mathcal I_t$:
  \begin{align} \label{eq:sos_constraint_xn_subsetneq_xs}
     -B_i \in \textbf M(V).
  \end{align}
}

\subsection{Resulting SOS problem}

The solution to \cref{prob:finding_cbfs_and_clf} -- that is the synthesis of suitable functions $B_i(x)$, $i \in \mathcal I_{n_{\!B}}$, and $V(x)$ -- can be obtained by solving the following SOS optimization problem:
\begin{align} \label{eq:sos_opt_prob}
   \hspace{-\leftmargin}
   \begin{array}{ll}
      \mbox{find} & V, B_i, p \in (\R[x])^m, s \in \Sigma[x], \\
      \mbox{maximize} & \text{vol}(\mathcal X_s) \\
      \mbox{subject to} & 
      \ifthenelse{\boolean{shortversion}} {\text{SOS constraints \cref{eq:sos_constr}} \\
      & B_i \in \textbf M(w_i) \quad \forall i \in \mathcal I_{n_{\!B}} \\
      & -B_i \in \textbf M(V) \quad \forall i \in \mathcal I_{n_{\!B}} \\
      }{
        \text{\cref{eq:sos_constr,eq:sos_constraint_xs_subsetneq_xa,eq:sos_constraint_xn_subsetneq_xs}} \\
      }
      & s - \epsilon \in \Sigma[x]. \\
   \end{array}
\end{align}
Here, the constraints encode set the CBF and CLF conditions, the containment properties \cref{eq:xs_subsetneq_xa,eq:xn_subsetneq_xs}, and a constant $\epsilon > 0$ assures that $s(x) > 0$.
\ifthenelse{\boolean{shortversion}} {
  Additional SOS constraints can be incorporated to capture linear and quadratic input constraint, as detailed in \cite{schneeberger2024advanced}.
} {
  Additional SOS constraints can be incorporated to capture linear and quadratic input constraint \cref{eq:linear_input_sos_constraints,eq:quadratic_input_sos_constraints}.

}
% where \cref{eq:sos_constr} encode the CBF and CLF condition, \cref{eq:linear_input_sos_constraints,eq:quadratic_input_sos_constraints} encode linear and quadratic input constraints, \cref{eq:sos_constraint_xs_subsetneq_xa,eq:sos_constraint_xn_subsetneq_xs} encode the set containment properties, and a constant $\epsilon > 0$ assures that $s(x) > 0$.

\subsection{Alternating Algorithm}

The solution of the SOS problem \cref{eq:sos_opt_prob} can be found by iteratively alternate between searching over one set of decision variables while keeping the others fixed, see \cite{TanPackard:04,schneeberger2023sos}.
In this process, either constraint margins are minimized or the volume of $\mathcal X_s$ is maximized.
The volume is approximated by the summation of traces of the gram matrix $Q_{B_i}$ in $B_i(x) = Z_{B,i}(x)^\top Q_{B,i} Z_{B,i}(x)$, similar to the approach in \cite{WangHan:18,KunduGeng:19}.
Such an alternating algorithm requires a feasible initialization of the decision variables that are kept fixed in the first iteration.
This is achieved through an initialization procedure that starts with an initially empty region of interest and expands it until it encompasses the allowable set $\mathcal X_a$.
\ifthenelse{\boolean{shortversion}} {
  For more details, we refer the reader to the supplementary material in \cite{schneeberger2024advanced}.
}{
  For more details, we refer the reader to the supplementary material of this paper.
}

\section{Safety Filter Implementation} \label{sec:online_controller}

% The objective of this section is to implement the advanced safety filter as specified in \cref{prob:finding_safety_filter}.
Given a linear input constraints set $\mathcal U$, we propose the following realization of the QP in closed loop for every state $x \in \R^n$ along the trajectory:
\begin{align} \label{eq:qcqp_based_controller}
   \begin{array}{llll}
      u_s(x) := &\underset{u \in \mathcal U}{\mbox{min}} & \| u_n(x) - u \|^2 \\
      &\mbox{s.t.} & C(x) u + b(x) \leq r(x), \\
   \end{array}
\end{align}
where
\begin{subequations} \label{eq:qcqp_based_controller_constraints}
   \begin{align}
      C(x) := \left[\begin{smallmatrix}
         \nabla V(x)^\top G(x) \\
         \nabla B_1(x)^\top G(x) \\
         \vdots \\
         \nabla B_{n_{\!B}}(x)^\top G(x) \\
      \end{smallmatrix} \right], \,
      b(x) := \left[ \begin{smallmatrix}
         \nabla V(x)^\top f(x) + d(x) \\
         \nabla B_1(x)^\top f(x) \\
         \vdots \\
         \nabla B_{n_{\!B}}(x)^\top f(x)
      \end{smallmatrix} \right]
   \end{align}
\end{subequations}
encode the CLF and CBF conditions \cref{eq:cbf_condition,eq:clf_condition}.
The smooth functions $r(x) = [\begin{matrix} r_1(x) & \hdots & r_{n_{\!B}}(x) \end{matrix}]$ represent state-dependent slack variables that specify the allowed dissipation rate and safety violation.
These slack-variables can be found using an SOS optimization, subject to upper and lower bounds that enforce the specifications outlined in \cref{prob:finding_safety_filter}.
% On the one hand, they are upper bounded to ensure the CLF and CBF conditions.
% On the other hand, they are lower bounded to ensure feasibility for the specific input $u=u_n(x)$ within $\mathcal X_n$ and for any $u \in \mathcal U$ over all $\mathcal X_s$.
% To ensure existence of a feasible $r(x)$, the upper bounds must be strictly greater than the lower bounds for all states.
% they are lower bounded to ensure feasibility for the specific input $u=u_n(x)$ within $\mathcal X_n$ and for any $u \in \mathcal U$ over all $\mathcal X_s$.

\begin{remark}
   \textit{
      In the literature on safety filters, see for example \cite{AmesCoogan:19}, the negative of the allowed dissipation rate and the safety violation are usually specified by class $\mathcal K$ functions $\alpha(V(x))$ and $\gamma_i(B_i(x))$:
      \begin{subequations}
         \begin{align}
            &\nabla V(x)^\top (f(x) + G(x) u) \leq -\alpha(V(x)), \\
            &\nabla B_i(x)^\top (f(x) + G(x) u) \leq -\gamma_i(B_i(x)) \quad i \in \mathcal I_{n_{\!B}}.
         \end{align}
      \end{subequations}
      In this case, the class $\mathcal K$ functions $\alpha$ and $\gamma_i$ depend on the value of $V(x)$ and $B_i(x)$.
      However, in \cref{eq:qcqp_based_controller_constraints}, we consider a broader class of functions $r_0(x)$ and $r_i(x)$, $i \in \mathcal I_{n_{\!B}}$, to ensure the preservation of the legacy control behavior during nominal conditions.
      % However, in \cref{eq:qcqp_based_controller_constraints}, we consider a broader class of the allowed dissipation rate violation $r_0(x)$ and allowed safety violation, $r_i(x)$, $i \in \mathcal I_{n_{\!B}}$, \afterReviewChangedBox{to ensure the preservation of the legacy control behavior during nominal conditions.}
      % the preservation of the legacy control action within the nominal region.
      % to ensure a smooth safety filter behavior, while satisfying further requirements presented below.
   }
\end{remark}

\subsection{Upper bounds on $r(x)$}

Within the domain of CLF and CBF conditions \cref{eq:cbf_condition,eq:clf_condition}, the slack-variables must satisfy the upper bounds, given by
\begin{align} \label{eq:r0_upper_bound}
   r_0(x) \leq 0
\end{align}
for all $x \in \mathcal X_t$, and
\begin{align} \label{eq:ri_upper_bound}
   r_i(x) \leq 0
\end{align}
for all $i \in \mathcal I_{n_{\!B}}$ and $x \in \partial \mathcal X_s$.

% To enforce the CLF condition \cref{eq:clf_condition}, $r_0(x)$ is upper bounded by:
% % On the one hand, the slack-variables are upper bounded to ensure
% % To enforce the CLF and CBF conditions \cref{eq:cbf_condition,eq:clf_condition} where required, the slack variables are upper bounded as:
% %  on the regions where they need to hold, that is
% \begin{align} \label{eq:r0_upper_bound}
%    r_0(x) \leq 0
% \end{align}
% for all $x \in \mathcal X_t$.
% To enforce the CBF condition \cref{eq:cbf_condition}, $r_i(x)$, $i \in \mathcal I_{n_{\!B}}$, are upper bounded by:
% \begin{align} \label{eq:ri_upper_bound}
%    r_i(x) \leq 0
% \end{align}
% for all $x \in \partial \mathcal X_s$.

\subsection{Lower bounds on $r(x)$}

To ensure tracking of the legacy controller $u_n(x)$ within $\mathcal X_n$ (cf. \cref{eq:sf_inactive_condition}), we impose the following lower bound on $r(x)$:
% On the other hand, they are lower bounded to ensure feasibility for the specific input $u=u_n(x)$ within $\mathcal X_n$ and for any $u \in \mathcal U$ over all $\mathcal X_s$.
% The first condition translates to the lower bound:
\begin{align} \label{eq:r_lower_bound_un_tracking}
   C(x) u_n(x) + b(x) \leq r(x)
\end{align}
for all $x \in \mathcal X_n$.
This condition guarantees that the cost in the QP in \cref{eq:qcqp_based_controller} can be minimized to zero within $\mathcal X_n$ by choosing $u = u_n(x)$, ensuring that the safety filter $u_s(x)$ coincides with the legacy control action $u_n(x) \in \mathcal U$.
The existence of $r_0(x)$ satisfying \cref{eq:r0_upper_bound,eq:r_lower_bound_un_tracking} is ensured through \cref{eq:forward_invariant_nominal_region}, while the existence of $r_i(x)$, $i \in \mathcal I_{n_{\!B}}$, satisfying \cref{eq:ri_upper_bound,eq:r_lower_bound_un_tracking} is ensured if $\mathcal X_n \subseteq \text{int}(\mathcal X_s$).

Furthermore, the selection of $r(x)$ must be lower-bounded to ensure the feasibility of the QP \cref{eq:qcqp_based_controller} within $\mathcal X_s$.
Although the exact lower bound is unknown, it can not be directly obtain as it involves an existence quantifier $\exists u \in \mathcal U$ for every state $x$.
However, it can be approximated using a feasible (but not necessarily optimal) solution $u_\text{sos}(x)$ from \cref{sec:sos_optimization}:
\begin{align} \label{eq:r_lower_bound_feasibility}
   C(x) u_\text{sos}(x) + b(x) \leq r(x)
\end{align}
for all $x \in \mathcal X_t$.
The existence of $r(x)$ satisfying \cref{eq:r0_upper_bound,eq:ri_upper_bound,eq:r_lower_bound_feasibility} is guaranteed through the SOS constraints \cref{eq:sos_constr_cbf,eq:sos_constr_clf}.
\ifthenelse{\boolean{shortversion}} {}{
  Typical lower and upper bound selections for $r_0(x)$ are illustrated in the supplementary material of this paper.
}

\subsection{Resulting SOS Problem}

The polynomial candidate $r(x)$ satisfying \cref{eq:r0_upper_bound,eq:ri_upper_bound,eq:r_lower_bound_un_tracking,eq:r_lower_bound_feasibility} can be obtained by solving the SOS problem:
\begin{align}
   \label{eq:sos_slack_variables}
   % \hspace{-\leftmargin}
   \begin{array}{ll@{}l}
      \mbox{find} & r \in (\R[x])^{t+1} \\
      \mbox{minimize} & \sum_i \text{tr}(Q_{r,i}) \\
      \mbox{subject to} & -r_0 &\in \textbf M(V, -B_{\mathcal I_{n_{\!B}}}) \\
      & -r_i &\in \textbf M(B_i, -B_{\mathcal I_{n_{\!B}}}) \\
      & s r_0 -\nabla V^\top \left ( s f + G p \right ) - s d \,\, &\in \textbf M(V, -B_{\mathcal I_{n_{\!B}}}) \\
      & s r_i -\nabla B_i^\top \left ( s f + G p \right ) &\in \textbf M(V, -B_{\mathcal I_{n_{\!B}}}) \\
      & r_0 -\nabla V^\top \left ( f + G u_n \right ) - d &\in \textbf M(-V) \\
      & r_i -\nabla B_i^\top \left ( f + G u_n \right ) &\in \textbf M(-V), \\
   \end{array}
\end{align}
where $B_i$, $V$, $p$, $s$ are taken from solving the SOS problem in \cref{eq:sos_opt_prob}, and the gram matrix $Q_{r,i}$ is defined by $r_i(x) = Z(x)^\top Q_{r,i} Z(x)$.
The cost function in \cref{eq:sos_slack_variables} favors choices of $r(x)$ that yield smooth intervention of the safety filter as the system approaches the state constraints.

\subsection{Lipschitz-continuity}

Finally, to establish local Lipschitz-continuity of the proposed safety filter, we need the following assumption.
\begin{assumption}[LICQ] \label{ass:licq}
   The Linear Independence Constraint Qualification (LICQ) is satisfied for the QP in \cref{eq:qcqp_based_controller_constraints} across all states $x$ and inputs $u$, that is, the gradients w.r.t. $u$ of the active inequality constraints in \cref{eq:qcqp_based_controller_constraints} are linearly independent.
   Equivalently, the rows of $C(x)$ associated with the active inequality constraints are linearly independent.
\end{assumption}
\begin{remark}
   \textit{
      By extending the QP in \cref{eq:qcqp_based_controller_constraints} with a slack variable $\delta \in \R^{t+1}$ as in \cite{ames2016control}
      \begin{align*}
         \begin{array}{llll}
            u_s(x) := &\underset{(u,\delta) \in \mathcal U \times \R^{t+1}}{\mbox{argmin}} & \| u_n(x) - u \|^2 + \lambda \| \delta \|^2 \\
            &\mbox{s.t.} & C(x)^\top u + b(x) \leq \delta, \\
         \end{array}
      \end{align*}
      we automatically satisfy the LICQ property, as demonstrated in \cite{ames2016control}.
      However, the introduction of these slack variables pose the potential risk of compromising safety and finite-time convergence guarantees.
      That being said, the LICQ assumption can easily be checked, and it generically holds true, i.e., for almost all polynomial coefficients, the Jacobian of a set of polynomials has full rank.
   }
\end{remark}

The local Lipschitz-continuity not only preserves existence and uniqueness of solutions (cf. \cite{Khalil:02}[Theorem 3.1]), but can also avoid the resulting feedback controller from chattering, as explained in \cite{morris2013sufficient}.
The following theorem establishes local Lipschitz-continuity of the developed QP controller.
\ifthenelse{\boolean{shortversion}} {
  Its proof can be found in the appendix of \cite{schneeberger2024advanced}.
} {
  Its proof is given in \cref{sec:proof_lipschitz}.
}
\begin{theorem} \label{thm:lipschitz_continuity}
   Under \cref{ass:licq},
   given a smooth $B(x)$ and $V(x)$ as specified in \cref{prob:finding_cbfs_and_clf},
   a smooth legacy controller $u_n(x)$,
   and twice differentiable $r_i(x)$, $i \in \{0\} \cup \mathcal I_{n_{\!B}}$, lower bounded by \cref{eq:r_lower_bound_feasibility},
   then $u_s(x)$ as defined in \cref{eq:qcqp_based_controller_constraints} is locally Lipschitz-continuous on the safe set $\mathcal X_s$.
\end{theorem}

% We can find polynomial candidates $r(x)$ based on given CBF $B(x)$ and CLF $V(x)$ by solving an additional SOS problem that encodes the upper and lower bounds \cref{eq:r0_upper_bound,eq:ri_upper_bound,eq:r_lower_bound_un_tracking,eq:r_lower_bound_feasibility}.

\section{Power Converter Case Study} \label{sec:simulations}

\begin{figure}[!t]
   \centering
   \resizebox{85mm}{!}{\includegraphics{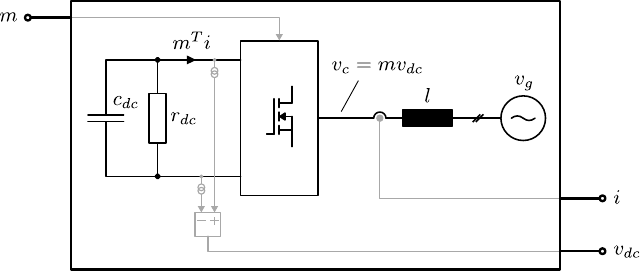}}
   \caption{
      The schematic is presented in the (dq) rotating frame, compromising a DC-link, 3-phase bridge, and an ac-line connected to an infinite bus $v_g$. Assuming the average-value model for the 3-phase bridge, the input $\bar u = \begin{bmatrix}m_d & m_q\end{bmatrix}^\top$ determines the converter voltage $v_c$ as a function of the dc-link voltage $v_{dc}$. Likewise, the dc-current flowing into the capacitor $c_{dc}$ is given by the input $\bar u$ and the ac-line current $i = \begin{bmatrix}i_d & i_q\end{bmatrix}^\top$. The states consisting of the dc-link voltage $v_{dc}$ and the line-current $i$ are measured.}
   \label{fig:power_converter_schematic}
\end{figure}

The proposed safety filter is implemented for a vector field realization of an ac/dc power converter model connected to an infinite bus, as illustrated in \cref{fig:power_converter_schematic}.
A feedback controller that avoids unsafe states is crucial for this application since all electrical variables (e.g. voltages and currents) need to be constrained at all times.
For more details about the modelling of power converters, we refer the reader to \cite{mohan2003power}.

The signals are represented in the (dq) rotating frame aligned with the phase $\theta = \omega_n t$ of the grid voltage $v_g$, where $\omega_n = 2 \pi 50$ is the nominal angular frequency in Hz.
The signals are given in per-unit representation.
The system's dynamics are given as follows:
\begin{subequations} \label{eq:power_converter_system_model}
\begin{align}
   (1/\omega_n) c_{dc} \dot v_{dc} &= - g_{dc} v_{dc} - i_d m_d - i_q m_q \\
   (1/\omega_n) l \dot i_d &= \omega l i_q + v_{dc} m_d - 1 \\
   (1/\omega_n) l \dot i_q &= -\omega l i_d + v_{dc} m_q,
\end{align}
\end{subequations}
where $l = 0.01$ is the inductance in p.u. consisting of transformer and line inductance, $c_{dc} = 0.02$ is the dc-link capacitance in p.u., and $g_{dc} := 1/r_{dc} = 0.001$ is the admittance of the DC-link capacitor in p.u.
The measured states are denoted by $\bar x := [\begin{matrix}v_{dc} & i_d & i_q\end{matrix}]^\top$, and the input is given by $\bar u := [\begin{matrix}m_d & m_q\end{matrix}]^\top$, describing the modulation indices.
The objective is to stabilize both the DC-link voltage $v_{dc}$ to $1$, and the quadrature current $i_q$ to its reference $i_{q,ref}$, which is assumed to be constant.
As a consequence, to meet the steady-state specifications, the direct current $i_d$ is stabilized to $-g_{dc}$.
The allowable set of states must adhere to the following inequality constraints:
\begin{align} \label{eq:power_converter_state_constraints}
   0.2 \leq v_{dc} \leq 1.2, \quad i_d^2 + i_q^2 \leq (1.2)^2.
\end{align}
% Furthermore, the quadratic input constraint $\mathcal U$ is given by
% \begin{align} \label{eq:power_converter_input_constraints}
%    m_d^2 + m_q^2 \leq (1.2)^2.
% \end{align}
The system's equilibrium is characterized by
\begin{align*}
   x^* = \begin{bmatrix}
      1 \\ -g_{dc} \\ i_{q,ref}
   \end{bmatrix},
   u^* = \begin{bmatrix}
      1 - \omega l i_{q,ref} \\ -\omega l g_{dc}
   \end{bmatrix}.
\end{align*}
We introduce the error coordinates for the states $x = \bar x - x^* = [\begin{matrix}\tilde v_{dc} & \tilde i_d & \tilde i_q\end{matrix}]^\top$ and the input variables $u = \bar u - u^* = [\begin{matrix}\tilde m_d & \tilde m_q\end{matrix}]^\top$ such that $\dot x = 0$ at $x=0$ and $u=0$.
The system represented in error coordinates is described by system $\dot x = f(x) + G(x) u$ with
\begin{subequations}
   \begin{align*}
      f(x) = \begin{bmatrix}
         -0.05 \tilde v_{dc} - 57.9 \tilde i_d + 0.00919 \tilde i_q \\
         1710 \tilde v_{dc} + 314 \tilde i_q \\
         -0.271 \tilde v_{dc} - 314 \tilde i_d
      \end{bmatrix}
   \end{align*}
   \begin{align*}
      G(x) = \begin{bmatrix}
         0.05 - 57.9 \tilde i_d & -57.9 \tilde i_q \\
         1710 + 1710 \tilde v_{dc} & 0 \\
         0 & 1710 + 1710 \tilde v_{dc}
      \end{bmatrix}.
   \end{align*}
\end{subequations}
We consider a legacy controller that resembles the proportional part of the synchronous rotating frame control structure, as described in \cite{blaabjerg2006overview}:
\begin{align*}
   u_n(x) = \begin{bmatrix}
      0.1 \tilde v_{dc} - \tilde i_d \\
      i_{q,ref} - \tilde i_q
   \end{bmatrix}.
\end{align*}
In the following subsection, we first consider a single quadrature current reference set to $i_{q,ref} = 0$.
Then, we introduce a non-zero quadrature current reference as a new state to the dynamical system.
% , with a zero time-derivative, i.e., ${\dot i}_{q,ref} = 0$.
% For each scenario, we find the CBFs $B_i(x)$ and CLF $V(x)$ (cf. \cref{prob:finding_cbfs_and_clf}), and perform simulations that illustrate the additional guarantees provided by the advanced safety filter $u_s(x)$ compared with a standard safety filter implementation.
% We use \texttt{CVXOPT} \textit{Python} library to solve the SDPs of the alternating algorithm, and to implement the QP in \cref{eq:qcqp_based_controller}.
% This library implements an optimization problem solver based on second-order and positive semidefinite cones.
% The simulations are performed using the ODE solver from \texttt{scipy} with a sampling time of $1$ $\mu s$.
The full code used for the optimization is available online\footnote{https://github.com/MichaelSchneeberger/advanced-safety-filter/}.

\begin{figure*}[!t]
   \centering
   \setkeys{Gin}{width=0.24\linewidth}
   \subfloat[\label{fig:stream_plot1}]{\resizebox{59mm}{!}{\includegraphics{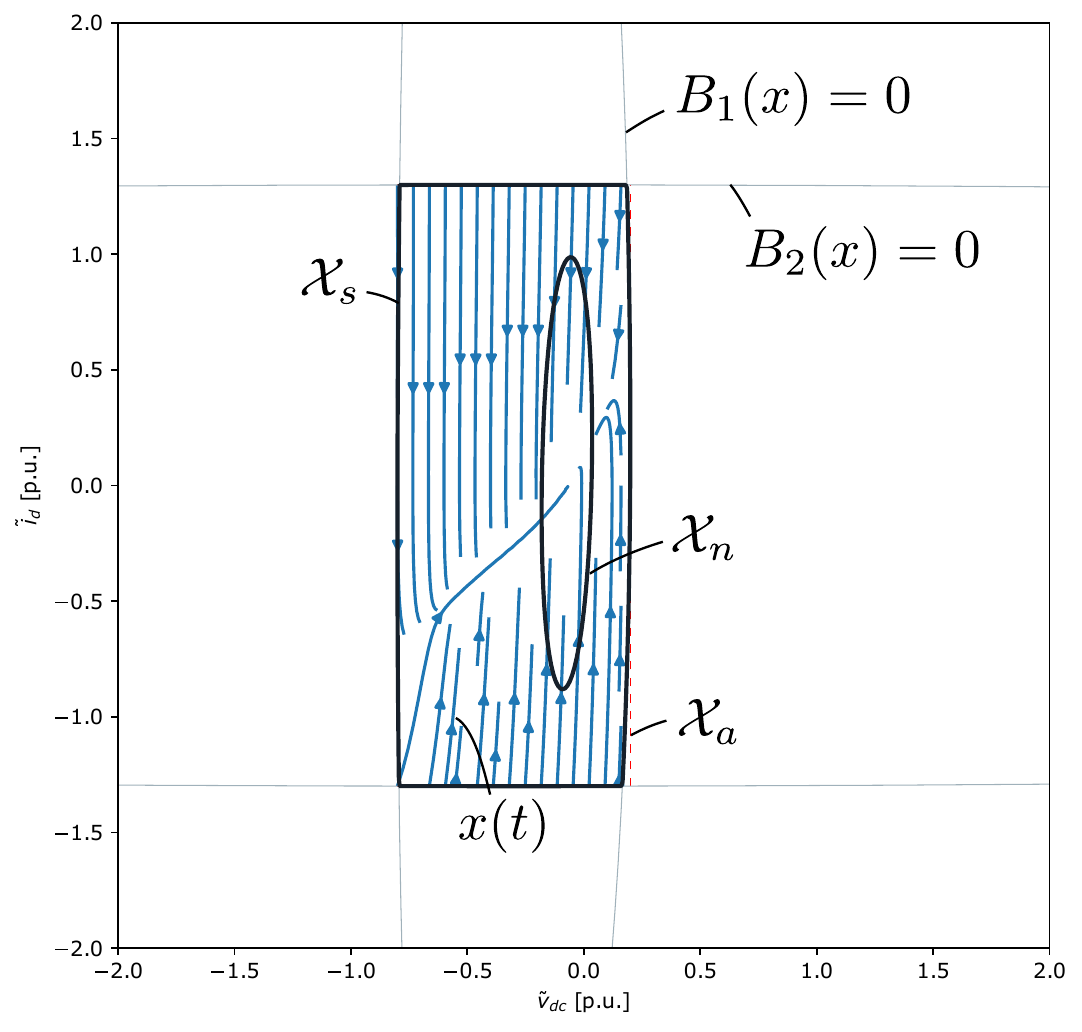}}}
   \hfil
   \subfloat[\label{fig:stream_plot2}]{\resizebox{59mm}{!}{\includegraphics{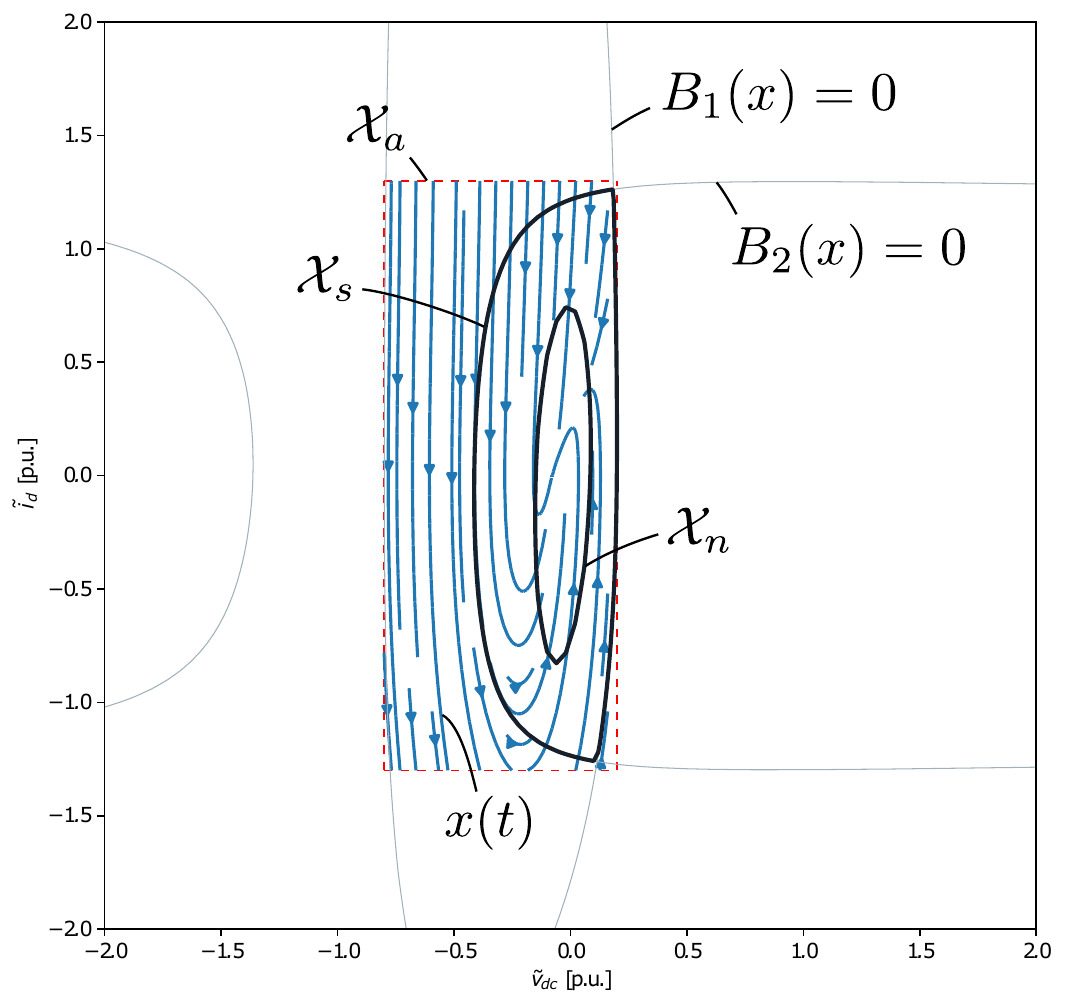}}}
   \hfil
   \subfloat[\label{fig:stream_plot3}]{\resizebox{59mm}{!}{\includegraphics{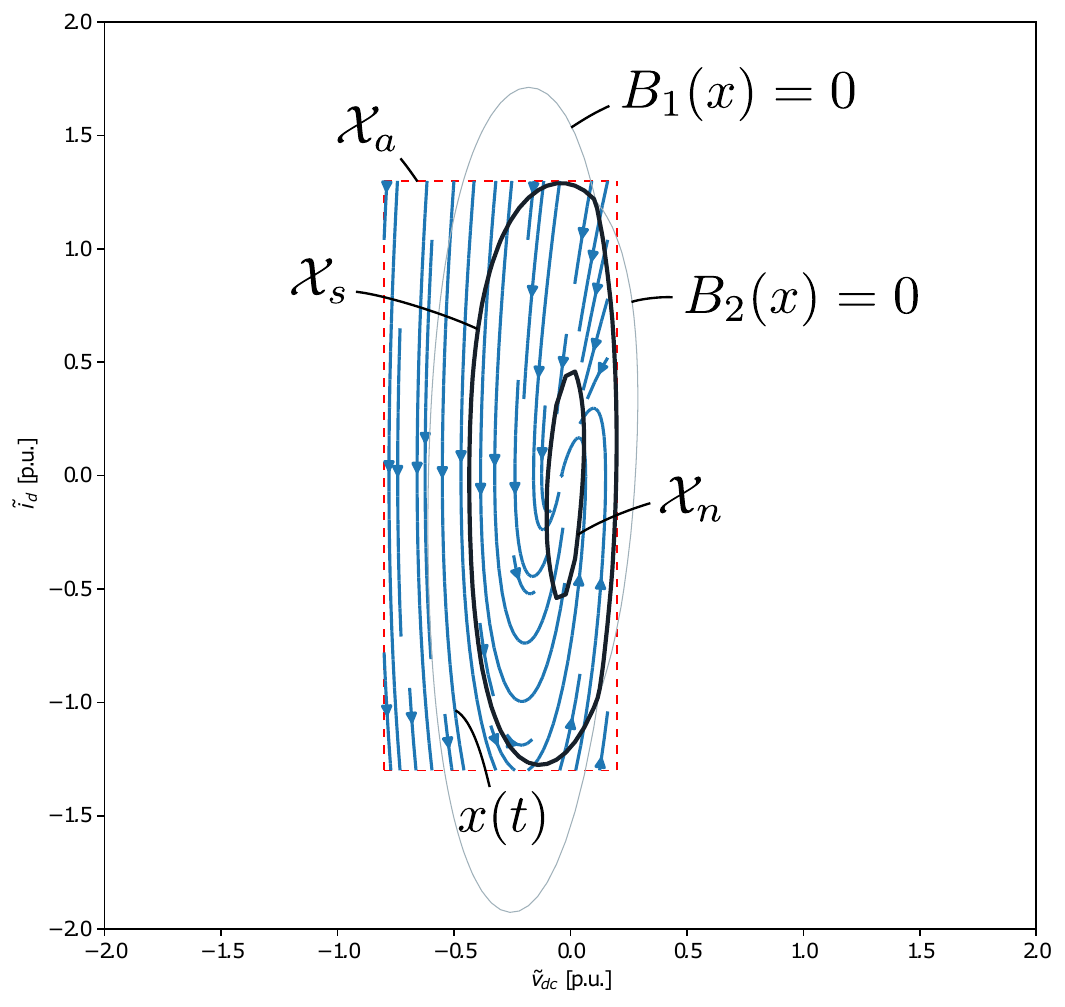}}}
   \caption{
      The allowable set $\mathcal X_a$, the safe set $\mathcal X_s$, and the nominal region $\mathcal X_n$ projected to the $\tilde v_{dc}$ and $\tilde i_d$ coordinate are shown for $\tilde i_q = i_{q,ref} = 0$. 
      In addition, the zero-level sets of the polynomials $B_1(x)$ and $B_2(x)$ are shown in gray.
      Three scenarios are shown, where (a) the input action is subjected to no constraints, i.e. $u(x(t)) \in \R^m$, (b) the input action is subjected to a quadratic constraint given by $u(x(t)) \in \mathcal U := \{ u \mid u^\top u \leq 1.2 \}$, and (c) $i_{q,ref}$ is integrated as a new state of the dynamical system using the same input constraints.
      The corresponding vector field $\dot x = f(x) + G(x) u_{sos}(x)$ (projected to $(\tilde v_{dc}, \tilde i_d)$ coordinates) is shown in blue, where $u_{sos}(x) = p(x)/s(x)$ is used to ensure compatibility between the CBF and CLF.
   }
   \label{fig:stream_plot}
\end{figure*}

\subsection{Zero quadrature current reference tracking}

In this subsection, we assume a zero quadrature current reference, i.e., $i_{q,ref} = 0$.
The allowable set $\mathcal X_a$, as expressed in \cref{eq:allowable_set}, encoding the state constraints \cref{eq:power_converter_state_constraints}, are achieved through the following polynomials:
\begin{subequations}
   \begin{align*}
      w_1(x) &= ((\tilde v_{dc} + 0.3)/0.5)^2 - 1 \\
      w_2(x) &= ({\tilde i_d}/1.3)^2 + ({\tilde i_q}/1.3)^2 - 1.
      % w_1(x) &= ((\tilde v_{dc} + 0.3)/0.5)^2 + (\tilde i_d/20)^2 + (\tilde i_q/20)^2 - 1 \\
      % w_2(x) &= (\tilde v_{dc}/20)^2 + ({\tilde i_d}/1.3)^2 + ({\tilde i_q}/1.3)^2 - 1.
   \end{align*}
\end{subequations}

Figures \ref{fig:stream_plot} (a) and (b) illustrates the safe set $\mathcal X_s$ and the nominal region $\mathcal X_n$ computed by the algorithm both without and with considering inputs constraints.
\cref{tab:number_of_var_and_constr} summarizes both the number of variables associated with the respective SDPs and the number of constraints, derived from the summation of triangular elements within the positive semidefinite matrices.
The initialization procedure completed in 10 iterations, where each iteration took around 10 seconds.
The main algorithm completed in 20 iterations, where each iteration took around 14 seconds.
The degree of the polynomials are given in \cref{tab:polynomial_degrees}.
\ifthenelse{\boolean{shortversion}} {}{
  Simulation results comparing the advanced safety filter with a basic safety filter are provided in the supplementary material.
}

\begin{table}[!t]
   \caption{Summary of SDP variables and constraints}
   \label{tab:number_of_var_and_constr}
   \centering
   \begin{tabular}{||cccccc||} 
      \hline
      Scenario & & init 1 & init 2 & main 1 & main 2 \\ \hline\hline
      Fig. \ref{fig:stream_plot}(a) & Variables & 475 & 187 & 410 & 475 \\
      & Constraints & 5726 & 4425 & 5807 & 5726 \\ \hline
      Fig. \ref{fig:stream_plot}(b) & Variables & 506 & 197 & 441 & 506 \\
      & Constraints & 6102 & 4765 & 6183 & 6102 \\ \hline
      Fig. \ref{fig:stream_plot}(c) & Variables & 1086 & 357 & 981 & 1086 \\
      & Constraints & 21277 & 17327 & 21474 & 21277 \\ \hline
   \end{tabular}
\end{table}
\begin{table}[!t]
   \caption{Degree of the polynomials encoding the CLF, CBF, and the rational controller}
   \label{tab:polynomial_degrees}
   \centering
   \begin{tabular}{||c@{\hskip 0.04in} c@{\hskip 0.04in} c@{\hskip 0.04in} c||} \hline
      $V(x)$ & $B_i(x)$ & $p(x)$ & $s(x)$ \\ \hline\hline
      4 & 4 & 3 & 2 \\ \hline
   \end{tabular}
\end{table}

\subsection{Including current reference tracking}

For this subsection, the objective is to stabilize both the DC-link voltage to $1$, and the quadrature current to some (not yet a priori pre-specified) quadrature current reference $i_{q,ref}$.
To render our design parametric in $i_{q,ref}$, we introduce a new state representing the quadrature current reference.
Consequently, we augment the dynamics from \cref{eq:power_converter_system_model} with $\dot i_{q,ref} = 0$.

\begin{remark}
   \textit{
      Although the reference $i_{q,ref}$ is assumed constant for the design of the CBFs and CLF, step-wise adjustments to $i_{q,ref}$ over time (corresponding to task cycles of the digital processing unit) do not compromise safety, provided the state remains within the safe set $\mathcal X_s$.
   }
\end{remark}

The corresponding coordinates of the states and inputs are given by $x = \begin{bmatrix}\tilde v_{dc} & \tilde i_d & \tilde i_q & \tilde i_{q,ref}\end{bmatrix}^\top$ and $u := \begin{bmatrix}\tilde m_d & \tilde m_q\end{bmatrix}^\top$, where $\tilde i_{q,ref} = i_{q,ref}$.
% The allowable set $\mathcal X_a$ is chosen through the following polynomials:
% \begin{align*}
%    w_1(x) &= \left ((\tilde v_{dc} + 0.3)/0.5 \right )^2 - 1 \\
%    w_2(x) &= ({\tilde i_d}/1.3)^2 + ({\tilde i_q}/1.3)^2 - 1.
%    % w_1(x) &= ((\tilde v_{dc} + 0.3)/0.5)^2 + (\tilde i_d/20)^2 + (\tilde i_q/20)^2 \\
%    % &+ (\tilde i_{q,ref}/20)^2 - 1 \\
%    % w_2(x) &= (\tilde v_{dc}/20)^2 + ({\tilde i_d}/1.3)^2 + ({\tilde i_q}/1.3)^2 \\
%    % &+ (\tilde i_{q,ref}/20)^2 - 1.
% \end{align*}

Correspondingly, a safe set $\mathcal X_s$ and the nominal region $\mathcal X_n$ can be computed for this 4-dimensional space.
\cref{fig:stream_plot}(c) illustrates the safe set $\mathcal X_s$ and the nominal region $\mathcal X_n$ for $i_q = i_{q,ref} = 0$.
\cref{tab:number_of_var_and_constr} summarizes the number of variables of the corresponding SDP for each part, as well as the number of constraints.
The initialization procedure completed in 51 iterations, where each iteration took around 60 seconds.
The main algorithm completed in 20 iterations, where each iteration took around 90 seconds.
% The degree of the polynomials are summarized in \cref{tab:polynomial_degrees}.

\section{Conclusion} \label{sec:conclusion}

We successfully developed and implemented a novel safety filter approach that, to our knowledge, is the first of its kind to leave the existing legacy controller subject to input constraints unaffected.
This innovation could lead to greater acceptance of safety filters in industrial environments and real-world scenarios, advancing their implementation across a broader range of applications.
Unlike traditional safety filters, which often require manual tuning and lack formal guarantees on preserving the nominal controller, our approach provides an automated and principled alternative that ensures interpretability and minimal intervention by design.
To support practical deployment, we also developed a software framework that automates the computation of the proposed safety filter.
\ifthenelse{\boolean{shortversion}} {
\afterReviewChangedBox{Future work will focus on a rigorous analysis of robustness to approximation errors in the legacy controller.}
} {
Future work will focus on a rigorous analysis of robustness to approximation errors in the legacy controller.
}
% \afterReviewChangedBox{Future work will focus on a rigorous analysis of robustness to approximation errors in the legacy controller.}
% Our next step is to integrate this safety filter with our industrial partners for the control of an advanced converter topology.

\ifthenelse{\boolean{shortversion}} {} {
  \section{Appendix}

  \subsection{Proof of \cref{thm:finite_time_convergence}} \label{sec:proof_finite_time_convergence}

  Because the safe set $\mathcal X_s$ is compact and forward invariant, and the closed-loop dynamical system $\dot x = f(x) + G(x) u_s(x)$ is Lipschitz-continuous, there exists a unique trajectory $x(t)$ for all $x(0) \in \mathcal X_s$ and $t > 0$.
  The CLF condition in \cref{eq:clf_condition} with $u := u_s(x)$ can be written without dissipation rate as $\forall x \in \partial X_n$:
  \begin{align} \label{eq:clf_condition_no_diss_rate}
     \nabla V(x)^\top \left ( f(x) + G(x) u_s(x) \right ) \leq 0.
  \end{align}
  Using the tangent cone $T_{\mathcal X_n}(x) = \{ z \mid \nabla V(x)^\top z \leq 0 \}$, the condition \cref{eq:clf_condition_no_diss_rate} can be formulated as $\forall x \in \partial X_n$:
  \begin{align*}
     f(x) + G(x) u_s(x) \in \mathcal T_{\mathcal X_n}(x).
  \end{align*}
  Hence, according to \cite[Nagumo's Theorem 4.7]{Blanchini:99}, $\mathcal X_n$ is forward invariant.
  This proves that for all $x(0) \in \mathcal X_n$, we can choose $T=0$.
  lt remains to show that for all other initial states, i.e., $x(0) \in \mathcal X_t$, there exists $T \geq 0$ such that $x(T) \in \mathcal X_n$, or equivalently $V \left ( x(T) \right ) \leq 0$.

  As $\mathcal X_t \subseteq \mathcal X_s$ is compact (cf. \cref{ass:compact_safe_set}) and the dissipative rate $d(x)$ defined in \cref{eq:dissipation_rate} is assumed to be strictly positive and Lipschitz-continuous on $\mathcal X_t$, the following minimum exists: $l = \underset{x \in \mathcal X_t}{\text{min}} \lambda(x) \left ( V(x) + V_0 \right ) > 0$.
  Therefore, the scalar function $V \left ( x(t) \right )$ strictly decreases over time for all $x(0) \in \mathcal X_t$ due to the CLF condition in \cref{eq:clf_condition}: $\dot V = \nabla V(x) \left (f(x) + G(x) u_s(x) \right ) \leq - l < 0$.
  As $\mathcal X_s$ is forward invariant, $V \left ( x(t) \right )$ is upper bounded by $V \left ( x(t) \right ) \leq V \left (x(0) \right ) - l t$  for all $x(0) \in \mathcal X_t$.
  Therefore, there exists some $T \leq V \left (x(0) \right ) /l$ such that $V \left (x(T) \right ) \leq 0$.

  \subsection{Derivation of the SOS constraints \cref{eq:sos_constr}} \label{sec:derive_sos_constr}

  The objective of this section is to convert the empty-set condition \cref{eq:empty-set_conditions_clf} into the SOS constraint given in \cref{eq:sos_constr_clf}. The derivation of \cref{eq:sos_constr_cbf} and \cref{eq:sos_constr_nom} is analogous.
  Applying Putinar's P-satz to \cref{eq:empty-set_conditions_clf} yields the SOS constraint
  \begin{align*}
     &-\nabla V^T f - d \in \textbf M(\nabla V^T G, -\nabla V^T G, V, -B_{\mathcal I_t}).
  \end{align*}
  By utilizing the definition of a quadratic module in \cref{eq:quadratic_module}, we can express this SOS constraint by the condition that there exist two SOS polynomials $p_{0,1}, p_{0,2} \in (\Sigma[x])^m$ such that
  \begin{align*}
     &-\nabla V^T \left ( f + G (p_{0,1} - p_{0,2}) \right ) - d \in \textbf M(V, -B_{\mathcal I_t}).
  \end{align*}
  By exploiting the fact that the set of SOS polynomials, when combined with the set of negative SOS polynomials, spans the entire set of polynomials (i.e., $\Sigma[x] - \Sigma[x] = R[x]$), we can reformulate the SOS constraint as the condition that there exists a polynomial $p_0 \in (R[x])^m$ such that
  \begin{align*}
     &-\nabla V^T \left ( f + G p_0 \right ) - d \in \textbf M(V, -B_{\mathcal I_t}).
  \end{align*}
  Finally, we introduce an additional SOS polynomial $s_0(x)$ (cf. \cref{eq:putinars_psatz_sos_condition}) and rewrite the SOS constraint as the condition that there exists $s_0 \in \Sigma[x], p_0 \in (R[x])^m$ such that
  \begin{align*}
     &-\nabla V^T \left ( s_0 f + G p_0 \right ) - s_0 d \in \textbf M(V, -B_{\mathcal I_t}).
  \end{align*}
  The introduction of the polynomial $s_0(x)$ increases the set of feasible solutions when truncating the degrees of the involved polynomials, as described in \cref{sec:putinars_psatz}.

  \subsection{Proof of \cref{lem:sos_constraint_sufficient}} \label{sec:proof_sos_constraint_sufficient}
  By dividing \cref{eq:sos_constraint_inequality_cbf,eq:sos_constraint_inequality_clf} by $s(x) > 0$, we obtain
  \begin{subequations}
     \begin{align}
        \begin{split} \label{eq:sos_constraint_inequality_cbf_div_by_s}
           \nabla B_i^\top \left ( f + G u_\text{sos} \right ) + \frac{\gamma_{0,i}}{s} B_i - \frac{\gamma_{1,i}^\top}{s} B_{\mathcal I_{n_{\!B}}} &\leq 0
        \end{split} \\
        \begin{split} \label{eq:sos_constraint_inequality_clf_div_by_s}
           \nabla V^\top \left ( f + G u_\text{sos} \right ) + d + \frac{\gamma_{0,0}}{s} V - \frac{\gamma_{1,0}^\top}{s} B_{\mathcal I_{n_{\!B}}} &\leq 0,
        \end{split}
     \end{align}
  \end{subequations}
  where $u_\text{sos}(x) = p(x)/s(x)$.

  From \cref{eq:sos_constraint_inequality_cbf_div_by_s}, by using the fact that $\gamma_{1,i}(x) \geq 0$, $s(x) > 0$, and $B_i(x) = 0$ and $B_{\mathcal I_{n_{\!B}}}(x) \leq 0$ on $\partial \mathcal X_s$, we infer, for each $i \in \mathcal I_{n_{\!B}}$, that for all $x \in \partial \mathcal X_s$ there exists $u := u_\text{sos}(x)$ such that $\nabla B_i^\top \left ( f + G u_\text{sos} \right ) \leq 0$
  whenever $B_i(x) = 0$.
  Or, equivalently, for all $x \in \partial \mathcal X_s$ there exists $u := u_\text{sos}(x)$ such that $\nabla B_i^\top \left ( f + G u \right ) \leq 0$
  for all $i \in \text{Act}(x)$, as defined in \cref{eq:cbf_condition_active_constraints}.
  This proofs the CBF condition in \cref{eq:cbf_condition} using the definition of the tangent cone under \cref{ass:regular_safe_set}.

  From \cref{eq:sos_constraint_inequality_clf_div_by_s}, by using the fact that $\gamma_{0,0}(x) \geq 0$, $\gamma_{1,0}(x) \geq 0$, $s(x) > 0$, and $V(x) \geq 0$ and $B_{\mathcal I_{n_{\!B}}}(x) \leq 0$ on $\mathcal X_t$, we deduce that for all $x \in \mathcal X_t$ there exists $u := u_\text{sos}(x)$ such that 
  $\nabla V^\top \left ( f + G u \right ) + \lambda \bigl ( V - V_0 \bigr ) \leq 0$.
  This proofs the CLF condition \cref{eq:clf_condition}.

  Finally, from \cref{eq:sos_constraint_inequality_nom}, by using the fact that $V(x) = 0$ on $\partial \mathcal X_n$, we deduce that the inequality $\nabla V^\top \left ( f + G u_n \right ) \leq 0$
  holds for all $x \in \partial \mathcal X_n$.
  This proofs the CBF condition in \cref{eq:forward_invariant_nominal_region}.

  \subsection{Proof of \cref{lem:quadratic_input_sos_constraints}} \label{sec:proof_quadratic_input}

  We use the definition of a quadratic module in \cref{eq:quadratic_module}, to rewrite the SOS constraint in \cref{eq:linear_input_sos_constraints} as $\forall x \in \R^n, \exists \gamma_u \in (\Sigma[x])^\top$:
  \begin{align*}
     -a_i u_\text{SOS}(x) + b_i + \frac{\gamma_u(x)^\top}{s(x)} B_{\mathcal I_{n_{\!B}}}(x) \geq 0
  \end{align*}
  for $i \in \mathcal I_l$, and $u_\text{SOS}(x) = p(x) / s(x)$.
  Since the $\gamma_u(x) \geq 0$, $s(x) > 0$, and $B_{\mathcal I_{n_{\!B}}}(x) \leq 0$ on $\mathcal X_s$, we deduce, for all $i \in \mathcal I_l$, $-a_i u_\text{SOS}(x) + b_i \geq 0$ on $\mathcal X_s$.
  This concludes the first part of the proof.

  By substituting the variable $v$ in \cref{eq:quadratic_input_sos_constraints} with $u - u_0$, we can reformulate the SOS constraints by using the definition of a quadratic module as $\forall x \in \R^n, \exists \gamma_{u,1} \in \Sigma[x, u], \gamma_{u,2} \in (\Sigma[x, u])^\top$:
  \begin{align} \label{_eq:quadratic_input_sos_constraints_div_by_s}
     \begin{split}
        &- \left ( (u - u_0)^\top Q_u \left ( u_\text{SOS}(x) - u_0 \right ) - u_\text{max}^2 \right ) \\
        &\quad + \frac{\gamma_{u,1}(x,u)}{s(x)} f_u(x, u - u_0) + \frac{\gamma_{u,2}(x,u)^\top}{s(x)} B_{\mathcal I_{n_{\!B}}} \geq 0,
     \end{split}
  \end{align}
  where $x \in \R^n$ and $u \in \R^m$ are the variables of the polynomial expressions.
  Because of its positive semidefiniteness, $Q_u$ can be decomposed into $Q_u = B_u^\top B_u$.
  Furthermore, since $\gamma_{u,1}(x,u) \geq 0$, $\gamma_{u,2}(x,u) \geq 0$, $s(x) > 0$, $B_{\mathcal I_{n_{\!B}}}(x) \leq 0$ on $\mathcal X_s$, and $f_u(x, u - u_0) \leq 0$ on $\mathcal U$, condition \cref{_eq:quadratic_input_sos_constraints_div_by_s} can be reformulated as $\forall x \in \mathcal X_s, \forall u \in \mathcal U$:
  \begin{align} \label{eq:quadratic_input_sos_constraints_local_on_xs}
     (u - u_0)^\top B_u^\top B_u (u_\text{SOS}(x) - u_0) \leq u_\text{max}^2.
  \end{align}

  For any $x \in \{ x \in \mathcal X_s \mid B_u (u_\text{SOS}(x) - u_0) = 0 \}$, we have $(u_\text{SOS}(x) - u_0)^\top B_u^\top B_u (u_\text{SOS}(x) - u_0) = 0 \leq u_\text{max}^2$.
  Hence, 
  $u_\text{SOS}(x) \in \mathcal U$.

  For any other safe state, that is $x \in \{ x \in \mathcal X_s \mid B_u (u_\text{SOS}(x) - u_0) \neq 0 \}$, there exists a unique $\lambda > 0$ such that 
  \begin{align} \label{eq:input_constr_lambda_condition}
     \begin{split}
        \left (B_u \lambda (u_\text{SOS}(x) - u_0) \right )^\top \left (B_u \lambda (u_\text{SOS}(x) - u_0) \right ) &= u_\text{max}^2.
     \end{split}
  \end{align}
  Additionally, substituting $u$ by $\lambda (u_\text{SOS}(x) - u_0) + u_0$ in condition \cref{eq:quadratic_input_sos_constraints_local_on_xs} yields the condition that $\forall x \in \mathcal X_s$:
  \begin{align} \label{eq:input_constr_lambda_condition_2}
     \lambda (B_u (u_\text{SOS}(x) - u_0))^\top (B_u (u_\text{SOS}(x) - u_0)) \leq u_\text{max}^2.
  \end{align}
  By comparing \cref{eq:input_constr_lambda_condition,eq:input_constr_lambda_condition_2}, we conclude that $\lambda$ must be greater than $1$.
  As a consequence, $(B_u (u_\text{SOS}(x) - u_0))^\top (B_u (u_\text{SOS}(x) - u_0)) \leq u_\text{max}^2$ holds for all $x \in \mathcal X_s$.
  This concludes that $u_\text{SOS}(x) \in \mathcal U$ for all $x \in \mathcal X_s$.

  \subsection{Proof of \cref{thm:lipschitz_continuity}} \label{sec:proof_lipschitz}

  Consider the optimization problem
  \begin{align} \label{eq:abstract_optimization_problem}
     \begin{array}{ll}
        \mbox{min}_u & f(u, x) \\
        \mbox{s.t.} & g(u, x) \leq 0,
     \end{array}
  \end{align}
  and a point $x^* \in \R^n$.
  We say $(u_s(x^*), \lambda(x^*))$ is a \emph{regular minimizer} if the LICQ is satisfied at $u_s(x^*)$, and $(u_s(x^*), \lambda(x^*))$ satisfies the KKT condition and the strong second order sufficient condition (SSOSC).
  The point $(u, \lambda)$ satisfies the SSOSC if Hessian matrix of the Lagrangian of \cref{eq:abstract_optimization_problem} is positive definite at this point.
  According to \cite[Theorem 2.3.3]{jittorntrum1978sequential}, there exists a locally Lipschitz continuous map $u_s(x)$ around a neighborhood of $x^*$ such that $u_s(x) \in \R^m$ is a minimizer for the non-linear optimization problem \cref{eq:abstract_optimization_problem}, if $f$ and $g$
  are twice continuously differentiable in $(u, x)$, and $u_s(x^*)$ is a regular minimizer with multiplier $\lambda(x^*)$.
  In what follows, we establish these conditions for the QP in \cref{eq:qcqp_based_controller_constraints}.

  LICQ property under \cref{ass:licq} asserts that for any solution of the QP $u_s(x^*)$, there exists a unique $\lambda^*$ that satisfies the KKT condition at $(u_s(x^*), \lambda^*)$.
  Next, we show that the SSOSC holds for any choice of $\lambda^*$, specifically $\lambda^*$ that satisfies the KKT condition.
  Consider the Lagrangian of the QP for either linear or quadratic input constraints:
  \begin{align*}
     \begin{split}
        L(u, \lambda) =& u^\top u + 2 u_n(x) u + u_n(x)^\top u_n(x)  \\
        & + \lambda_1^\top ( C(x) u + b(x) ) \\
        & + \lambda_2^\top ( u^\top Q_u u - u_{max}^2 ) + \lambda_3^\top (A u - b)
     \end{split}
  \end{align*}
  Then the inequality $ v^\top \nabla_{uu}^2 L(u, \lambda) v = v^\top v + \lambda_2 v^\top Q_u v > 0$ for all $v \in \R^m$.
  This proves that any solution of the QP $u_s(x^*)$ satisfies the SSOSC with the corresponding $\lambda^*$.
  Furthermore, a solution of the QP $u_s(x)$ exists for all $x$ due to the lower bound on $r(x)$ in \cref{eq:r_lower_bound_feasibility}, consequently establishing the local Lipschitz continuity of $u_s(x)$.
}

\bibliographystyle{unsrt}
\bibliography{main}

\vskip -2\baselineskip plus -1fil

% supplementary material
\ifthenelse{\boolean{shortversion}} {}{
  \section{Supplementary Material}
\label{sec:supplementary_material}

\subsection{Alternating Algorithm} \label{sec:algorithm}

The solution of the SOS problem \cref{eq:sos_opt_prob} presents a particular challenge due to its bilinearity, i.e., the simultaneous search for CBF $B(x)$, CLF $V(x)$, a rational controller $u_\text{SOS}(x)$, and other SOS polynomials introduced by the Positivstellensatz.
This bilinearity prevents us from converting the SOS problem directly into a Semi-Definite Program (SDP).
Similar to \cite{TanPackard:04,schneeberger2023sos}, however, we can iteratively alternate between searching over one set of decision variables while keeping the others fixed.
Such an alternating algorithm requires a feasible initialization of the decision variables that are kept fixed in the SDP of the first iteration.
Therefore, we propose an initialization procedure to find such an initial set of feasible parameters following the presentation of the main algorithm.

\subsection{Main algorithm} \label{sec:main_algorithm}

To make the main algorithm compatible with the initialization procedure, we introduce an \emph{operating region} that contains the allowable set $\mathcal X_a$:
\begin{align} \label{eq:radius_of_interest}
   \mathcal X_\text{op} := \{ x \mid f_\text{op}(x) \leq 0 \} \supseteq \mathcal X_a,
\end{align}
with scalar polynomial $f_\text{op} \in R[x]$.
We assume that there exists a constant $\rho_\Sigma \in \R$ such that 
\begin{align} \label{eq:rho_sigma}
   f_\text{op}(x) + \rho_\Sigma \in \Sigma[x]
\end{align}
is an SOS polynomial.
This assumption, while not overly restrictive, is important for establishing the feasibility of the SDP in \textit{Part 1} of the initialization procedure, as elaborated in the subsequent subsection.

\begin{remark}
   \textit{
      The operating region $\mathcal X_\text{op}$ can be determined by solving another SOS problem, as condition \cref{eq:rho_sigma} directly translates to an SOS constraint.
      By making minor adjustments, it is possible to define the operating region $\mathcal X_\text{op}$ using multiple polynomials, and it could even coincide with the allowable set $\mathcal X_a$.
      However, in order to reduce computational complexity, we restrict the number of polynomials to one.
      In cases, where the operating region $\mathcal X_\text{op}$ closely approximates the safe set $\mathcal X_s$, the polynomials $B_i(x)$ in the quadratic modules of the SOS problem in \cref{eq:sos_opt_prob} can be omitted, thereby reducing the total number of decision variables.
      This will reduce computational complexity by introducing slightly more conservative SOS constraints.
   }
\end{remark}

The constraint $f_\text{op}(x) \leq 0$ defining the operating region $\mathcal X_\text{op}$ is added to the empty set conditions \cref{eq:empty-set_conditions} developed in \cref{sec:sos_optimization}.
From a set-theoretic point of view, adding the constraint $f_\text{op}(x) \leq 0$ does not alter the empty set conditions, given that $\mathcal X_s \subseteq \mathcal X_\text{op}$.
However, by applying Putinar's P-satz on the empty set conditions \cref{eq:empty-set_conditions} including $f_\text{op}(x) \leq 0$, we obtain slightly different SOS constraints for all $i \in \mathcal I_t$:
\begin{subequations} \label{eq:sos_constraint_roi}
   \begin{align}
      \begin{split}
         &-\nabla V^T \left ( s f + G p \right ) - s d \in \textbf M(V, -B_{\mathcal I_t}, -f_\text{op})
      \end{split} \\
      \begin{split}
         &-\nabla B_i^T \left ( s f + G p \right ) \in \textbf M(B_i, -B_{\mathcal I_t}, -f_\text{op})  
      \end{split} \\
      \begin{split}
         &-\nabla V^T \left ( f + G u_n \right ) - d \in \textbf M(V, -V, -f_\text{op}).
      \end{split}
   \end{align}
\end{subequations}
These SOS constraints are used to define the main algorithm.

The main algorithm involves alternating between two SDPs, as described by \textit{Part 1} and \textit{Part 2} below.
Given an initial set of feasible parameters, the SDP in \textit{Part 1} is guaranteed to be feasible for each iteration when utilizing the solution from \textit{Part 2}.
Likewise, the SDP in \textit{Part 2} is guaranteed to be feasible for each iteration when utilizing the solution from \textit{Part 1}.
Furthermore, as shown in \cite{schneeberger2023sos}, the cost of the SDP in \textit{Part 1} encoding the volume of the safe set is non-decreasing over the iterations.

\subsubsection*{Part 1 - Searching for CBF and CLF}

In the first part, we search over CBFs $B_i(x)$ and a CLF $V(x)$ that maximize the volume of the safe set.
To make the SDP computationally tractable, we approximate the volume of the safe set by the summation of traces of $Q_{B_i}$, similar to \cite{WangHan:18,KunduGeng:19}:
\begin{align*} 
   \hspace{-\leftmargin}
   \begin{array}{lll}
      \mbox{find} & V, B_i \in R[x] \\
      \mbox{hold fixed} & p, s \\
      \mbox{maximize} & \sum_{i \in \mathcal I_t} tr(Q_{B_i}) \\
      \mbox{subject to} & \text{\cref{eq:sos_constraint_xs_subsetneq_xa,eq:sos_constraint_xn_subsetneq_xs,eq:sos_constraint_roi}}, \\
   \end{array}
\end{align*}
where the $Q_{B_i}$ are derived from the decomposition of the CBF $B_i(x)$ into $B_i(x) = Z_{B,i}(x)^T Q_{B,i} Z_{B,i}(x)$ (cf. \cref{eq:quadratic_matrix_form}).

\subsubsection*{Part 2 - Searching for a controller}

For the second part of the main algorithm, we search over a rational controller $u_\text{SOS}(x) = p(x)/s(x)$ that maximizes the SOS constraint margins, denoted by $\Delta_i \in \R$, $i \in \{ 0 \} \cup \mathcal I_t$.
Only the first two SOS constraints from \cref{eq:sos_constraint_roi} depend on the rational controller.
They are specified with the inclusion of these margins $\Delta_i$ as:
\begin{subequations} \label{eq:sos_constraint_margin}
   \begin{align}
      \begin{split}
         &-\nabla V^T \bigl ( s f + G p \bigr ) - s d + \Delta_0 \in \textbf M(B_i, -B_{\mathcal I_t}, -f_\text{op}) \\
      \end{split} \\
      \begin{split}
         &-\nabla B_i^T \bigl ( s f + G p \bigr ) + \Delta_i \in \textbf M(B_i, -B_{\mathcal I_t}, -f_\text{op})
      \end{split}
   \end{align}
\end{subequations}
for all $i \in \mathcal I_t$.
Large values of $\Delta_i$, $i \in \{ 0 \} \cup \mathcal I_t$, will increase the set of feasible solution in \textit{Part 1} in the subsequent iterations.
The SDP involved in \textit{Part 2} is given by:
\begin{align} \label{eq:sos_opt_prob_part2}
   \hspace{-\leftmargin}
   \begin{array}{lll}
      \mbox{find} & p \in (\R[x])^m, s \in \Sigma[x], \Delta_i \in \R \\
      \mbox{hold fixed} & V, B_i \\
      \mbox{maximize} & \sum_{i \in \{0\} \cup \mathcal I_t} \Delta_i \\
      \mbox{subject to} & \text{\cref{eq:linear_input_sos_constraints,eq:quadratic_input_constraints,eq:sos_constraint_margin}} \\
      & s - \epsilon \in \Sigma[x], \Delta_\text{min} \leq \Delta_i.
   \end{array}
\end{align}
A constant $\epsilon > 0$ assures that $s(x) > 0$.
The minimum margin $\Delta_\text{min} < 0$ is selected to maintain numerical stability of the algorithm.
The main algorithm concludes when the increment in $\sum_{i \in \mathcal I_t} tr(Q_{B_i})$ in \textit{Part 1} does not exceed a predefined minimal threshold.

\subsection{Initialization procedure} \label{sec:init_procedure}

\begin{figure}[!t]
   \centering
   \resizebox{85mm}{!}{\includegraphics{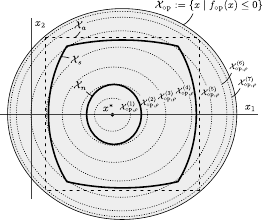}}
   \caption{(Operating region):
      The initialization procedure produces an initial set of feasible parameters for the main algorithm by iteratively expanding the operating region $\mathcal X_{\text{op},\rho}$.
      The initialization procedure finishes 
      if $\mathcal X_\text{op} = \mathcal X_{\text{op},\rho}^{(k)}$.
   }
   \label{fig:growing_region_of_interest}
\end{figure}

The main algorithm, presented in previous subsection, requires an initial set of feasible parameters.
Therefore, we present an initialization procedure to find such an initial set.

Similar to the main algorithm, the initialization procedure involves an alternating between two SDPs, as described by \textit{Part 1} and \textit{Part 2} below.
The feasibility of each part is guaranteed by employing the solution of the SDP from the previous iteration.
To make sure that the SDP in \textit{Part 1} is feasible in the first iteration, we use a modified operating region (cf. \cref{eq:radius_of_interest})
\begin{align} \label{eq:small_radius_of_interest}
   \mathcal X_{\text{op}, \rho} := \{ x \mid f_\text{op}(x) + \rho \leq 0 \},
\end{align}
with $\rho \geq 0$.
By choosing the $\mathcal X_{\text{op}, \rho}$ initially sufficiently small in volume, i.e., a large $\rho > 0$, we guarantee feasibility of the SDP in \textit{Part 1} for any initial set of parameters, as shown in \cref{lem:feasible_solution}.

By applying Putinar's P-satz on the modified empty set condition \cref{eq:empty-set_conditions} including $f_\text{op} + \rho \leq 0$ instead, we derive the following SOS constraints for all $i \in \mathcal I_t$:
\begin{subequations} \label{eq:sos_constraint_init_roi}
   \begin{align}
      \begin{split}
         &-\nabla V^T \left ( s f + G p \right ) - s d \in \textbf M(V, -B_{\mathcal I_t}, f_\text{op} + \rho)
      \end{split} \\
      \begin{split}
         &-\nabla B_i^T \left ( s f + G p \right ) \in \textbf M(B_i, -B_{\mathcal I_t}, f_\text{op} + \rho)  
      \end{split} \\
      \begin{split}
         &-\nabla V^T \left ( f + G u_n \right ) - d \in \textbf M(V, -V, f_\text{op} + \rho).
      \end{split}
   \end{align}
\end{subequations}

\subsubsection*{Part 1 - Searching for CBF and CLF}

The first part of the initialization procedure searches over CBF and a CLF that maximize $\rho$, thereby increasing the volume of $\mathcal X_{\text{op}, \rho}$ for every iteration (cf. \cref{fig:growing_region_of_interest}).
To make sure that $\mathcal X_{\text{op}, \rho}$ is contained in the operating region, i.e., $\mathcal X_\rho^{(k)} \subseteq \mathcal X_\text{op}$, we impose the constraint that $0 \leq \rho$.
The resulting SDP is expressed as:
\begin{align} \label{eq:sos_opt_prob_part1}
   \hspace{-\leftmargin}
   \begin{array}{lll}
      \mbox{find} & V, B_i \in R[x], \rho \in \R \\
      \mbox{hold fixed} & p, s \\
      \mbox{maximize} & \rho \geq 0 \\
      \mbox{subject to} & \text{\cref{eq:sos_constraint_xs_subsetneq_xa,eq:sos_constraint_xn_subsetneq_xs,eq:sos_constraint_init_roi}}. \\
   \end{array}
\end{align}

The next lemma establishes feasibility of the SDP \cref{eq:sos_opt_prob_part1} in the first iteration.
% Its proof can be found in the appendix.
\begin{lemma} \label{lem:feasible_solution}
   Assuming that there exist candidates $V(x)$ and $B_i(x)$ that satisfy conditions \cref{eq:sos_constraint_xs_subsetneq_xa,eq:sos_constraint_xn_subsetneq_xs}, then the SDP in \cref{eq:sos_opt_prob_part1} is feasible using random initial coefficient values for polynomials $p(x) \in (R[x])^m$ and $s(x) \in \Sigma[x]$.
\end{lemma}
\begin{IEEEproof}
   By utilizing the definition of a quadratic module in \cref{eq:quadratic_module}, the conditions in \cref{eq:sos_constraint_init_roi} can be formulated as the existence of $\gamma_{0,i}, \gamma_n \in R[x]$, $\gamma_{0,0}, \gamma_{\text{op}, 0}, \gamma_{\text{op},i}, \gamma_{\text{op},n} \in \Sigma[x]$, and $\gamma_{1,i}, \gamma_{1,0} \in (\Sigma[x])^t$ such that
   \begin{subequations} \label{eq:sos_constraint_lemma_4}
      \begin{align}
         \begin{split}
            & q_i + \gamma_{\text{op}, i} (f_\text{op} + \rho) \in \Sigma[x]   
         \end{split} \\
         \begin{split}
            & q_0 + \gamma_{\text{op}, 0} (f_\text{op} + \rho) \in \Sigma[x]
         \end{split} \\
         \begin{split}
            & q_n + \gamma_{\text{op}, n} (f_\text{op} + \rho) \in \Sigma[x],
         \end{split}
      \end{align}
   \end{subequations}
   with polynomials \begin{itemize}
      \item $q_i := -\nabla B_i^T \left ( s f + G p \right ) - \gamma_{0,i} B_i + \gamma_{1,i}^T B_{\mathcal I_t}$
      \item $q_0 := -\nabla V^T \left ( s f + G p \right ) - s d - \gamma_{0,0} V + \gamma_{1,0}^T B_{\mathcal I_t}$
      \item $q_n := -\nabla V^T \left ( f + G u_n \right ) - d + \gamma_n V$.
   \end{itemize}
   
   First, we show that for any polynomial $q \in R[x]$, we can find $\gamma_{\text{op}} \in \Sigma[x]$ and $\rho \in \R$ such that
   \begin{align} \label{eq:abstract_q_sos}
      q(x) + \gamma_{\text{op}}(x) (f_\text{op}(x) + \rho ) \in \Sigma[x].
   \end{align}
   To show this, we select an SOS polynomial $\gamma_{\text{op}} \in \Sigma[x]$ sufficiently large such that $q(x) + \gamma_{\text{op}}(x) \in \Sigma[x]$.
   This is always possible since starting from the decomposition $q(x) = Z(x)^T Q_f Z(x)$ (cf. \cref{eq:quadratic_matrix_form}), we can select $\gamma_{\text{op}}(x) = \nu Z(x)^T Z(x)$, with arbitrary large $\nu > 0$, such that $Q_f + \nu I \succcurlyeq 0$.
   Second, by utilizing the assumption on $\rho_\Sigma$ (cf. \cref{eq:rho_sigma}), choose $\rho := \rho_\Sigma + 1$ such that
   $q(x) + \gamma_{\text{op}}(x) + \gamma_{\text{op}}(x) (\rho(x) + \rho_\Sigma ) \in \Sigma[x]$.
   Hence, any $\rho \geq \rho_\Sigma + 1$ renders \cref{eq:abstract_q_sos} an SOS polynomial.
   
   In this way, we can find $\rho$ for each SOS constraint in \cref{eq:sos_constraint_lemma_4}, and denote them by $\rho_0, \rho_i, \rho_n \in \R$.
   By selecting $\rho := max(\rho_0, \rho_i, \rho_n)$, all SOS constraints in \cref{eq:sos_constraint_lemma_4} hold simultaneously.
   Hence, the SOS problem \cref{eq:sos_opt_prob_part1} is feasible for any initialization of $q_0(x)$, $q_i(x)$, $i \in \mathcal I_t$, and $q_n(x)$.
\end{IEEEproof}

\subsubsection*{Part 2 - Searching for a controller}

Similar to the main algorithm, we search over a rational controller $u_\text{SOS}(x) = p(x)/s(x)$ that maximizes the SOS constraint margins, denoted by $\Delta_i \in \R$, $i \in \{ 0 \} \cup \mathcal I_t$.
The corresponding SOS constraints including these margins $\Delta_i$ are specified as:
\begin{subequations} \label{eq:sos_constraint_init_margin}
   \begin{align}
      \begin{split}
         &-\nabla V^T \left ( s f + G p \right ) - d + \Delta_0 \in \textbf M(V, -B_{\mathcal I_t}, f_\text{op} + \rho)
      \end{split} \\
      \begin{split}
         &-\nabla B_i^T \left ( s f + G p \right ) + \Delta_i \in \textbf M(B_i, -B_{\mathcal I_t}, f_\text{op} + \rho),
      \end{split}
   \end{align}
\end{subequations}
for all $i \in \mathcal I_t$.
The resulting SDP is given by:
\begin{align*}
   \hspace{-\leftmargin}
   \begin{array}{lll}
      \mbox{find} & p \in (\R[x])^m, s \in \Sigma[x], \Delta_i \in \R \\
      \mbox{hold fixed} & V, B_i, \rho \\
      \mbox{maximize} & \sum_{i \in \{0\} \cup \mathcal I_t} \Delta_i \\
      \mbox{subject to} & \text{\cref{eq:linear_input_sos_constraints,eq:quadratic_input_constraints,eq:sos_constraint_init_margin}}, \\
      & s - \epsilon \in \Sigma[x], \Delta_\text{min} \leq \Delta_i.
   \end{array}
\end{align*}
A constant $\epsilon > 0$ assures that $s(x) > 0$.
Similar to the SDP \cref{eq:sos_opt_prob_part2}, the minimum margin $\Delta_\text{min} < 0$ is selected to maintain numerical stability of the algorithm.
The initialization procedure is considered complete when $\rho = 0$, i.e., $\mathcal X_\rho^{(k)} = \mathcal X_\text{op}$.
At this point, we found an initial set of feasible parameters for the SDP in \textit{Part 1} of the main algorithm.
\begin{remark}
   \textit{
      The initialization procedure can be extended with additional steps where the margins $\Delta_i$ are introduced when searching over the controller, or the operational region is expanded -- by decreasing $\rho$ -- when searching for CBF and CLF. 
      Furthermore, we can introduce additional steps where we optimize only over a subset of the $\Delta_i$'s.
   }
\end{remark}

\subsection{Lower and upper bounds of $r_0(x)$}

   \begin{figure}[!t]
      \centering
      \resizebox{80mm}{!}{\includegraphics{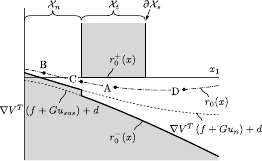}}
      \caption{(Upper and lower bounds on $r_0(x)$):
         Whereas the upper bound $r_0^+(x)$ is directly given by the CLF condition in \cref{eq:clf_condition}, only an estimate of the lower bound $r_0^-(x)$ can be obtained by the solution of the SOS constraints in \cref{eq:sos_opt_prob} involving $u_\text{sos} := p(x) / s(x)$ and the legacy controller $u_n(x)$.
      }
      \label{fig:bounds_on_ri}
   \end{figure}

   We illustrate a typical design choice for $r_0(x)$, as defined in \cref{eq:qcqp_based_controller}, in \cref{fig:bounds_on_ri}.

   For states inside the transitional region $x \in \mathcal X_t$ (see point \textbf{A} in \cref{fig:bounds_on_ri}), $r_0(x)$ is required to be smaller than the upper bound $r_0^+(x) = 0$ for \cref{eq:qcqp_based_controller_constraints} to satisfy the CLF condition in \cref{eq:clf_condition}.
   On the other hand, an estimate of the lower bound on $r_0(x)$ can be obtained from the solution $u_\text{sos}(x) = p(x)/s(x)$ of the SOS problem in \cref{eq:sos_opt_prob}.
   The corresponding lower bound is then given by $r_0^-(x) = \nabla V(x)^T \left ( f(x) + G(x) u_\text{sos}(x) \right ) + d(x)$, which is smaller than the upper bound $r_0^+(x) = 0$ within $\mathcal X_t$ due to \cref{eq:sos_constraint_inequality_clf}.
   We wish to emphasize that there may exist more optimal lower bounds on $r_0(x)$; however, the one chosen is certified by the SOS constraints developed in the previous chapter.

   For states inside the nominal region $x \in \mathcal X_n$ (see point \textbf{B} in \cref{fig:bounds_on_ri}), no upper bound is specified.
   Hence, by selecting a large $r_0(x)$, we can relax the corresponding constraint of the QCQP in \cref{eq:qcqp_based_controller_constraints}.
   Applying the same approach to other $r_i(x)$ allows us to enlarge the feasible set until $u := u_n(x)$ is a feasible solution for the QCQP, required to satisfy \cref{eq:sf_inactive_condition}.
   Therefore, we propose the following lower bound: $r_0^-(x) = \nabla V(x)^T \left ( f(x) + G(x) u_{n}(x) \right ) + d(x)$.
   By assigning $u := u_n(x)$, the corresponding constraint of the QCQP, expressed as $r_0^-(x) - r_0(x) \leq 0$, is clearly satisfied.

   For states on the boundary of the nominal region $x \in \partial \mathcal X_n$ (see point \textbf{C} in \cref{fig:bounds_on_ri}), we must ensure compatibility of the lower and upper bounds discussed in the previous paragraphs.
   More concretely, we need to ensure that $\nabla V(x)^T \left ( f(x) + G(x) u_{n}(x) \right ) + d(x) \leq 0$ for all $x \in \partial \mathcal X_n$.
   However, this inequality follows directly from inequality \cref{eq:sos_constraint_inequality_nom}.

   Finally, for states outside the safe set $x \notin \mathcal X_s$ (see point \textbf{D} in \cref{fig:bounds_on_ri}), an upper bound on $r(x)$ is not specified.
   The lower bound can be chosen as in \textbf{A}.
   It is however important to note that the region can be made attracting with respect to the safe set by the additional requirement of selecting a negative $r(x)$.
   This can restore the system state to the safe set in scenarios where unexpected system noise or model mismatch cause the system to deviate from the safe set.

\subsection*{Comparison with a Basic Safety Filter}

The proposed safety filter introduces additional guarantees, which make it more stringent and, as noted in the introduction, more suitable in already established industrial setups. 
However, due these system requirements, it now represents a new category that prioritizes the preservation of the legacy control, enhancing composability, preventing a direct performance comparison between conventional safety filters and our novel setup.
Nonetheless, to provide qualitative insight, side-by-side plots in \cref{fig:simulation_convergence_rate} are included that illustrate the additional guarantees compared to a basic safety filter in selected scenarios.
The advanced safety filter $u_s(x)$ is based on the previously computed CLF $V(x)$ and CBFs $B_1(x)$ and $B_2(x)$, and is compared to a basic safety filter $u_\text{SF}(x)$ using the same CBFs:
\begin{align} \label{eq:basic_safety_filter}
   \begin{array}{lll}
      \underset{u \in \mathcal U}{\mbox{argmin}} & \| u_n(x) - u \|^2 \\
      \mbox{s.t.} & \nabla B_1(x) (f(x) + G(x) u) + \alpha \left ( B_1(x) \right ) \leq 0 \\
      & \nabla B_2(x) (f(x) + G(x) u) + \alpha \left ( B_2(x) \right ) \leq 0,
   \end{array}
\end{align}
where $\alpha(B_i(x)) := \alpha B_i(x)$ and $\alpha = 10$ is selected.
\cref{fig:simulation_convergence_rate} illustrates how the advanced safety filter ensures legacy system behavior, while the basic safety filter suboptimally perturbs the legacy control action, resulting in a much slower convergence.
We note, however, that increasing $\alpha$ to 100 removes this effect: the basic safety filter now also closely follows the legacy system behavior.
\begin{remark}
   \textit{
In general, choosing a large $\alpha$ can reduce the likelihood of intervention and thereby preserve the legacy control input inside the safe set.
However, this heuristic approach does not provide the same a priori guarantees as our proposed method and may introduce undesirable closed-loop behavior. %, particularly near the boundary of the safe set.
  In particular, it can create undesired equilibrias near the boundary of the safe set \cite{reis2020control}, arising from the safety filter projecting an unsafe legacy-induced state derivative onto a zero state derivative vector.
In contrast, the proposed advanced safety filter guarantees convergence towards the nominal region with a prescribed dissipation rate $d(x)$.
% Furthermore, for a large $\alpha$, the safety filter becomes more sensitive on system mismatch or measurement noise when approaching the boundary of the safe set.
% This increased sensitivity can potentially result in closed-loop instabilities.
   }
\end{remark}

% \afterReviewAddedBox{While the basic safety filter can be manually tuned for improved performance, this approach lacks any a priori guarantees.}
% , as discussed in \cref{sec:safe_control_extension}.
% By increasing the value $\alpha$, it becomes less likely that the safety filter modifies the legacy control action within the safe set.
% On the other hand, for a large $\alpha$, the safety filter becomes more sensitive on system mismatch or measurement noise when approaching the boundary of the safe set.
% This increased sensitivity can potentially result in closed-loop instabilities.

\begin{figure}[!t]
   \centering
   \includegraphics[width=86mm]{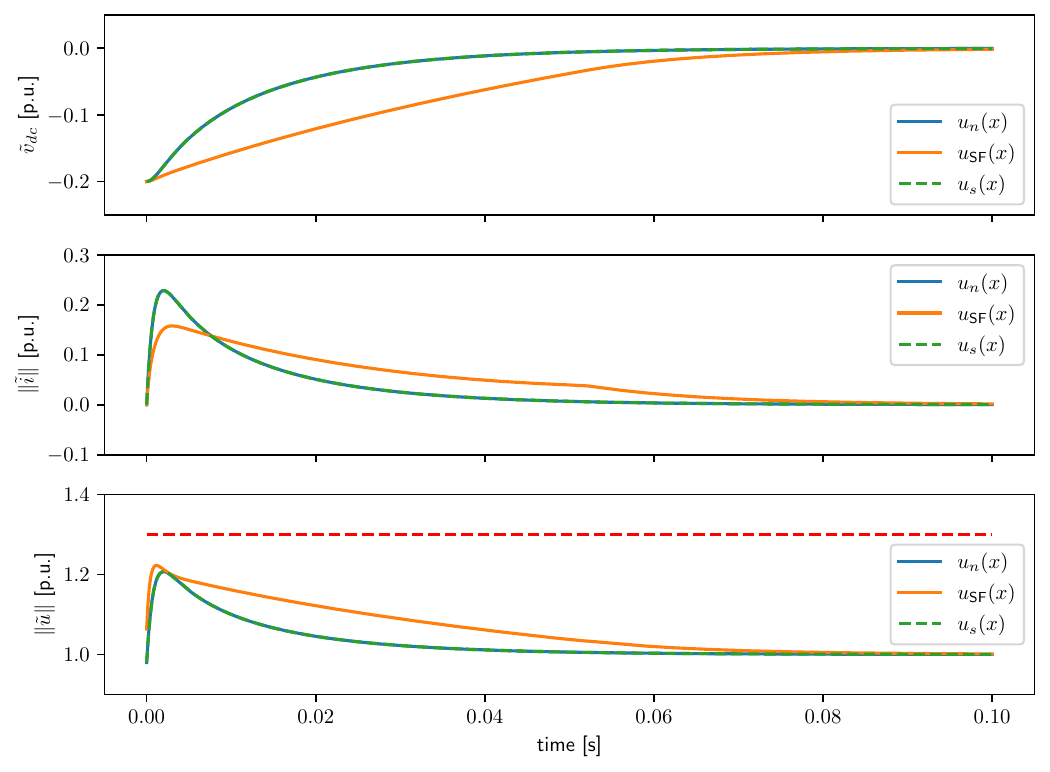}
   \caption{(Convergence to the nominal region $\mathcal X_n$):
      While the advanced safety filter $u_s(x)$ essentially overlaps the legacy control $u_n$,
      the basic safety filter $u_\text{SF}(x)$ suboptimally perturbs the legacy control action, resulting in a much slower convergence.
   }
   \label{fig:simulation_convergence_rate}
\end{figure}

A side-by-side plots in \cref{fig:simulation_constraint_violation} illustrate the effect on including quadratic input constraints into the design of the safety filter.
We consider the integration of quadratic input constraints into the SOS constraint as a contribution distinct from the discussion on the safety filter's role in preserving the legacy control action. 
However, for consistency, the simulations were performed using the previous setup, which including an advanced safety filter and the basic safety filter in \cref{eq:basic_safety_filter}.
Since the input constraints are incorporated into the design of the CLF and CBFs, the safety filter $u_s(x)$ successfully stays within the state constraints for all $t \geq 0$.
This is not true for a basic safety filter $u_\text{SF}(x)$, which does not consider the input constraints in its design.
In this case, the norm of the input action $\| u_\text{SF}(x) \|$ exceeds the input constraint limit of $1.3$ p.u. in the beginning of the simulation (cf. \cref{fig:simulation_constraint_violation}).
To produce a valid input, the input action $u_\text{SF}(x)$ is projected to the set of feasible input set $\mathcal U$.
As a result, the trajectory deviates from its intended path, leading to a violation of the dc-link voltage constraint.

\begin{figure}[!t]
   \centering
   \includegraphics[width=86mm]{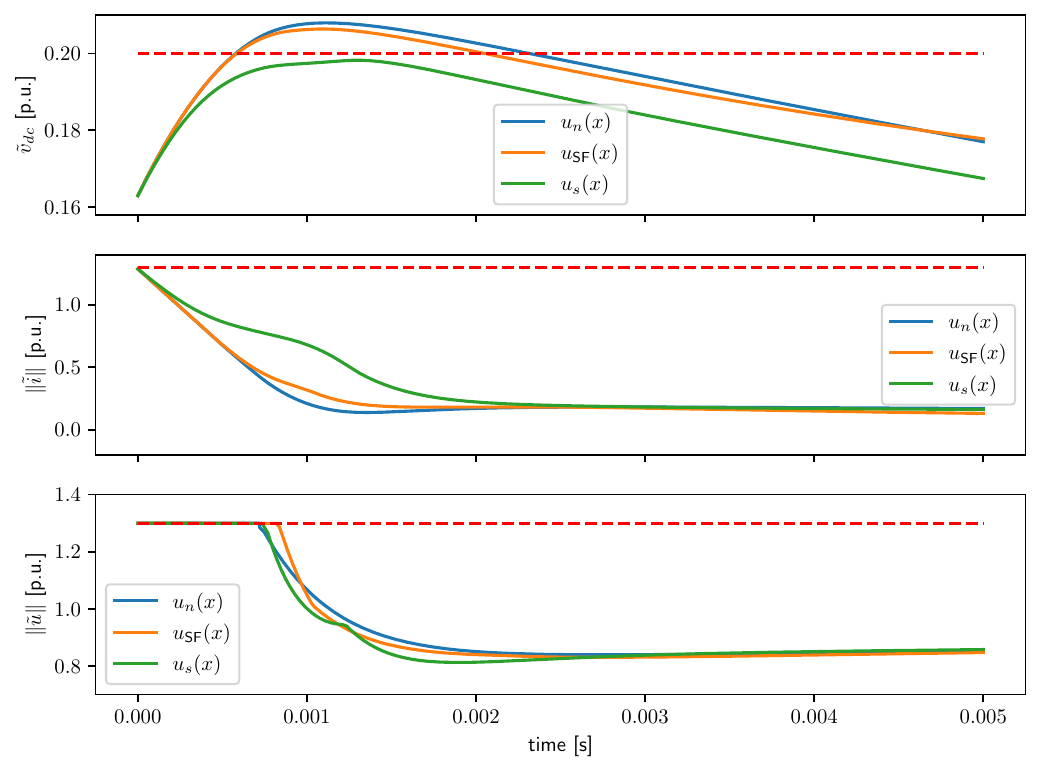}
   \caption{(Forward invariance of the safe set $\mathcal X_s$): 
      While the advanced safety filter $u_s(x)$, which considers the quadratic input constraints into the design, successfully stays within the state constraints, whereas the legacy controller $u_n$ and basic safety filter $u_\text{SF}(x)$ violate the state constraints.
   }
   \label{fig:simulation_constraint_violation}
\end{figure}

The simulation shown in \cref{fig:simulation_iqref_tracking} illustrates the reference tracking performance of our safety filter $u_s(x)$ compared to the basic safety filter $u_\text{SF}(x)$ introduced in the previous subsection.
The safety filter, although slightly perturbing the legacy controller, guarantees finite-time convergence to the nominal region defined for a quadrature current reference $i_{q,ref}$ of $1$.
However, the basic safety filter $u_\text{SF}(x)$ perturbs the legacy control action, leading to a much slower convergence.
In general, as we approach the boundary of the safe set by increasing $i_{q,ref}$, the basic safety filter is more likely to intervene the legacy control action.

\begin{figure}[!t]
   \centering
   \includegraphics[width=86mm]{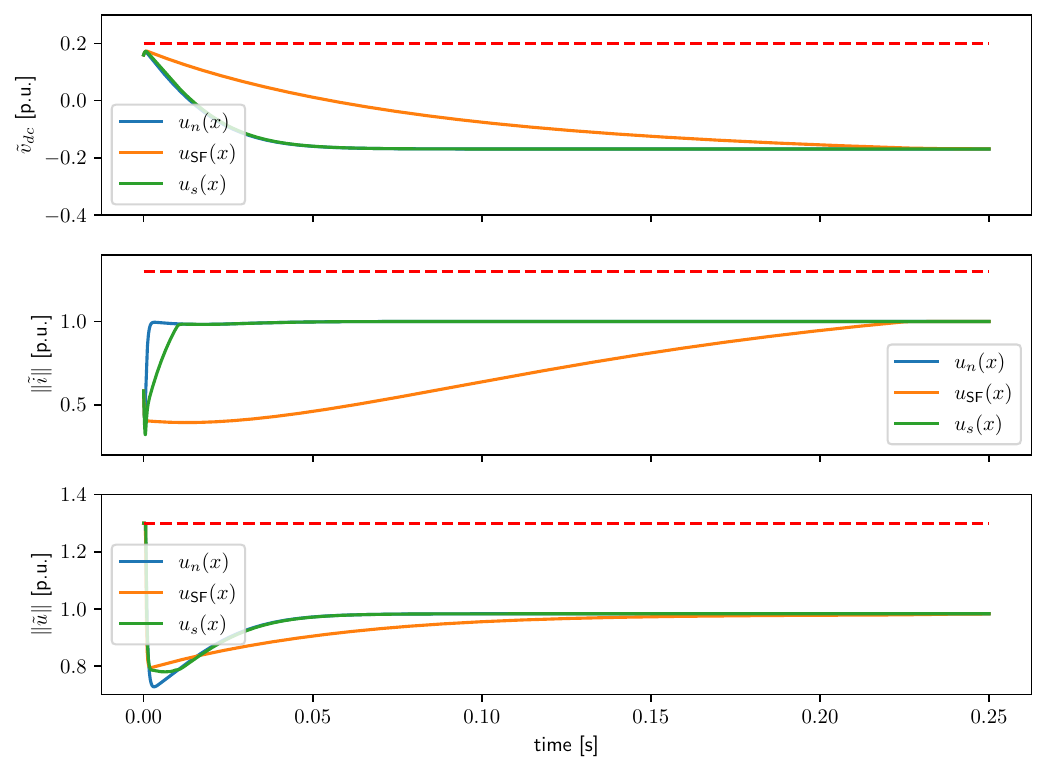}
   \caption{(Quadrature current reference tracking): 
      The closed-loop simulation demonstrates the finite-time convergence guarantee of the advanced safety filter $u_s(x)$ towards the nominal region $\mathcal X_n$ for non-zero quadrature current reference $i_{q,ref} = 1$p.u.
      Additionally, the advanced safety filter exhibits a faster convergence to the equilibrium compared to the basic safety filter $u_\text{SF}(x)$.
      The latter is more likely to disturb the legacy control action as the quadrature current reference $i_{q,ref}$ approaches the boundary of the safe set, resulting in a slower convergence.
   }
   \label{fig:simulation_iqref_tracking}
\end{figure}

}

\end{document}